\documentclass[useAMS,usenatbib,usenatNishiNishibib,usegraphicx]{mn2e}

\newcommand{\dpr}{$^{\prime\prime}$}
\newcommand{\pr}{$^{\prime}$}
\newcommand{\hi}{H~{\sc i} }
\newcommand{\hii}{H~{\sc ii} }

\newcommand{\lsun}{$L_\odot$}

\newcommand{\MB}{$M_{B}$}
\newcommand{\LB}{$L_{B}$}
\newcommand{\sqas}{arcsec$^{-2}$}

\newcommand{\dg}{$^{\circ}$}
\newcommand{\hr}{$^{h}$}
\newcommand{\mn}{$^{m}$}
\newcommand{\seg}{$^{s}$}

\title[Oxygen Gradients and the Corotation Radius]{Breaks in the Radial Oxygen Abundance and Corotation Radius of Three Spiral Galaxies}

\author[Scarano Jr. et al.]{S. Scarano Jr $^{1,2}$ \thanks{E-mail: scarano@astro.iag.usp.br /
scaranojr@ctio.noao.edu}, J. R. D. L\'{e}pine $^1$, M. M. Marcon-Uchida $^1$\\
$^1$ Instituto de Astronomia, Geof\'{\i}sica e Ci\^encias Atmosf\'ericas da
Universidade de S\~ao Paulo,\\ Cidade Universit\'aria, CEP: 05508-900, S\~ao Paulo, SP, Brazil \\
$^2$ Southern Astrophysical Research Telescope (SOAR), Casilla 603, La Serena,
Chile
\\}

\begin{document}
\maketitle

\begin{abstract}

The correlation between the breaks in the metallicity distribution
and the corotation radius of spiral galaxies has been already
advocated in the past and is predicted by a chemo-dynamical model of
our Galaxy that effectively introduces the role of spiral arms in
the star formation rate. In this work we present photometric and
spectroscopic observations made with the Gemini Telescope for three
of the best candidates of spiral galaxies to have the corotation
inside the optical disk: IC0167, NGC1042 and NGC6907. We observed
the most intense and well distributed \hii regions of these
galaxies, deriving reliable galactocentric distances and Oxygen
abundances by applying different statistical methods. From these
results we confirm the presence of variations in the gradients of
metallicity of these galaxies that are possibly correlated with the
corotation resonance.

\end{abstract}


\section{Introduction}

The question of what triggers the star formation in the disks of
galaxies, and in particular the  role of the spiral arms, has been
the subject of a long debate.  Soon after the theory of spiral arms
was proposed by \cite*{LinShu69}, \cite{Roberts69} showed that shock
waves are produced in the passage of the gas across the spiral
gravitational perturbation, and this could trigger star formation.
This was not a surprising result, since the arms are obvious to us
because they are the areas containing bright stars. \cite{Oort74},
analyzing the gas distribution in M51, suggested that a constant
fraction of the gas per passage through the arms is transformed into
stars, and consequently that it would be natural to expect the star
formation rate to be proportional to $\Omega-\Omega_p$.
\cite*{Jensen76} tested this expression with a sample of nearby
galaxies, by plotting a chemical enrichment indicator ([O~{\sc
iii}]/H$\beta$) as a function of $\Omega-\Omega_p$ , both quantities
being measured at a normalized radius, and concluded that the star
formation rate is proportional to $\Omega-\Omega_p$. However,
\cite{McCall86} argued that the correlation found by Jensen et al.
could be the result of independent processes having similar
dependence on galactic radius, and presented observations of {\it O}
abundance gradients in two flocculent galaxies, NGC5055 and NGC7793,
which were found to be very similar to those of grand design
spirals. As the flocculent galaxies were supposed to be free of
spiral arms, the arms could not be the cause of the observed
gradients.  This argument was later weakened by the fact that a
clear spiral structure was found  in NGC5055, both in the infrared
\citep{Thornley96} and in  molecular gas \citep{Kuno97}. Thornley
found infrared spiral structures in several other flocculent
galaxies, which is an indication that it is a normal characteristic
of these galaxies. Nevertheless, this does not prove that the spiral
arms are responsible for the observed gradients. The nature of the
flocculent galaxies is not fully understood, and it is not clear if
they can be considered (or not) as good examples of galaxies where
stochastic self-propagating star formation dominates. As a first
step, our effort will be to understand the star formation mechanism
in normal spiral galaxies. \cite*{Zaritsky92} investigated the {\it
O} abundance profile of seven nearby galaxies and found a distinct
change in slope for 3 of them. These changes cannot be associated
with calibration problems, since precisely the same methods were
used in all the cases. Later \cite*{Zaritsky94} extended the study
to a larger number of galaxies, and found for a few of them not only
a change of slope, but even possible reversals, with metallicity
increasing at large radii (see for instance the profiles of
NGC3369). It should be noted that a reversal is easily explained if
the star formation rate is considered to be proportional to
$|\Omega-\Omega_p|$, which is the same expression above but takes
into account the cases in which corotation is inside the zone being
investigated; a minimum of star formation is expected at
$\Omega=\Omega_p$. An interesting case is the Milky Way. A simple
model of chemical enrichment was proposed by Mishurov et al. (2002)
for our Galaxy, in which the star formation rate is proportional to
$\Omega-\Omega_p$. These authors argue that the minimum (or the
plateau) in the gradient of metallicity of our Galaxy, observed
using different tracers (eg. Maciel \& Quireza 1999, using planetary
nebulae and Andrievsky et al. 2004 using Cepheids) is coincident
with the corotation radius. The above interpretation for the
existence of minima or plateaus in the metallicity gradients is not
easy to verify in other galaxies, however, a correlation between
radii of minima and/or inflexions in the metallicity distribution
and the corotation radius was found by \citep{Sca2010} using
published data. One of the difficulties to identify the corotation
is that this resonance is not necessarily situated in the optical
disk. Nevertheless, \cite{Canzian98} proposed a test to identify
galaxies for which the corotation is favorably placed in the optical
disk. This test only requires moderately deep imaging and no
spectroscopic observations are needed. In this paper we present the
study of the radial gradients of metallicity in three of the best
candidates identified by the Cazian's test (IC0167, NGC1042 and
NGC6907) to verify if the connection between metallicity breaks and
corotation holds for these galaxies. The observations were performed
using the Gemini Multi-Object Spectrograph (GMOS), pointed towards
232 \hii regions distributed in these galaxies. Different
statistical methods were employed to get {\it O} abundances from the
observed emission lines. The sample of galaxies and \hii regions are
presented in the section 2 of this paper. Observations and data
reduction are described in section 3. The photometric and
spectroscopic results are presented in section 4. Finally the
results are discussed in section 5, where the corotation radii of
these galaxies are estimated.

\section{Sample Selection}

\subsection{Galaxies}

In order to perform our study we searched in the literature for
galaxies for which the corotation is expected to be favorably placed
in the optical disk. This was done by selecting three of the best
candidates tested by \cite{Canzian98} (Table \ref{tbl-1}) according
to the following criteria:

\begin{enumerate}

\item Galaxies should fail in the conservative version of the
Canzian's test, evaluated by the ratio of the outer to inner extent
of the spiral structure. In principle, this guarantees that the
corotation is situated in the optical part of the disk
\citep{Canzian98};

\item Objects with larger values of the ratio mentioned above were
preferred in the selection, since larger ratios means that the
corotation radius is deeper into the bright part of the galactic
disk \citep{Canzian98};

\item Only galaxies with angular sizes that fit entirely inside the
GMOS field (5.5 $\times$ 5.5 arcmin) were chosen. The purpose of
this was to cover all the extension of the galaxies with maximum
spatial resolution;

\item Each galaxy should have at least 16 \hii regions, since this is
the minimum number of \hii regions needed for an ideal sampling to
measure the metallicity gradient, according to \cite{DutilRoy01}.

\item The three brightest galaxies among those selected with the
previous criteria were chosen for observation.

\end{enumerate}

\begin{table}
 \caption{Parameters of the selected galaxies. $RA$ and $Dec$ are
the equatorial coordinates, $ratio$ is the measured ratio of the
outer to inner extent of the spiral structure, $V_{sys}$ is the
systemic velocity, $D$ is the Hubble flow distance for $V_{sys}$
($H_{0} = 73 km/s/Mpc$), $s$ is the linear scale in the plane of the
sky at distance $D$, $a$ and $b$ are respectively the sizes of the
major and minor axes, $m$ the total magnitude and $Type$ is the
morphological type. (1) Canzian (1998); (2) NED; (3) Scarano Jr et
al. (2008) }
 \label{tbl-1}

\begin{tabular}{llll}
 \hline
{\it {\bf Parameter}} & {\it {\bf IC0167}} & {\it {\bf NGC1042}} & {\it {\bf NGC6907}} \\
 \hline
{\it  RA}         & 1\hr51\mn08.5\seg & 2\hr40\mn24\seg  & 20\hr25\mn06.6\seg \\
{\it  Dec}        & 21\dg54\pr46\dpr  & -8\dg26\pr01\dpr & -24\dg48\pr34\dpr \\
{\it $ratio^{(1)}$}      & 2.41              & 2.38 / 3.31      & 2.85 \\
{\it $V_{sys}$ [km/s]}   & 2931$^{(2)}$      & 1371$^{(2)}$     & 3182$^{(3)}$ \\
{\it $D$ [Mpc]}          & 41.6              & 18.6             & 43.6 \\
{\it $s$ [pc/arcsec]}    & 202               & 90               & 211 \\
{\it $a^{(2)}$ [arcmin]} & 2.9               & 4.7              & 3.3 \\
{\it $b^{(2)}$ [arcmin]} & 1.9               & 3.6              & 2.7 \\
{\it $m^{(2)}$}          & 13.6              & 11.5             & 11.9 \\
{\it $Type^{(2)}$}       &  SAB(s)c          & SAB(rs)cd        & SB(s)bc \\
 \hline
\end{tabular}
\end{table}

The fact that the selected galaxies do not have strong bars may
simplify the interpretation of the data, since  only the secular
effect of the spiral arms may be predominant on these galaxies.

\subsection{H~{\sc ii} Regions}

To perform the spectroscopy we adopted the following procedures to
select the \hii regions of each galaxy :

\begin{enumerate}

\item Before the GMOS observations we elaborated a list of \hii
regions candidates based on the Second Palomar Sky Survey,
photometric calibrated using the second release of the USNO CCD
Astrograph Catalog (UCAC). The objects were selected according to
their flux profile, brightness and geometry;

\item Taking into account that the GMOS supports only one slit per
dispersion position and one fixed position angle per observation, we
determined instrumental position angles (Table \ref{tbl-2}) in order
to maximize the number of \hii regions candidates to be observed and
minimize the number of overlapping spectra in the acquisition
process. These position angles were evaluated using the code
PAMASKGEM
\footnote{http://www.ctio.noao.edu/ftp/pub/scarano/codes/pamaskgem.pro};

\item After the GMOS pre-imaging using the position angle obtained
in (ii) we selected the best candidates \hii regions in each
dispersion position. We chose the objects according to their
brightness, emission profile (to avoid foreground stars) and colors
(as a secondary criterium to exclude foreground stars). Special care
was taken to avoid the objects positioned in places where the
dispersed spectrum would generate spectral lines away from the GMOS
detectors.

\end{enumerate}

\section{Gemini observations}

\par The observations were conducted at the Gemini-North telescope
in the years 2005 and 2006 using the GMOS Multi-Slit Spectrograph in
queue mode. All the observations were executed in better conditions
than the minimum required (sky background of $50\%$, cloud cover of
$70\%$ and image quality of $70\%$), ensuring full spectrophotometric
conditions of the data.

\par
\begin{figure}
\includegraphics[width=84mm]{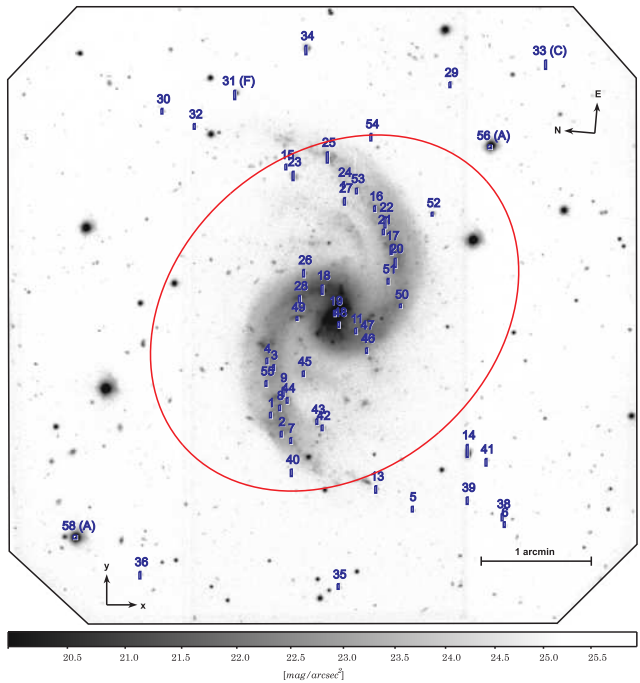} \caption{Slits superimposed to
the r\pr image of the galaxy IC0167. Only the slits used in the 2006
observations are presented. "A" are the stars used for alignment,
"F" are reference stars for flux calibration and "C" is a slit used
for sky sampling. The sky orientation is given by the N-E compass
and the spectral dispersion follows the x direction of the x-y
compass. The ellipse represents the projection of a circle in the
sky plane considering the same inclination and position angles
obtained from the spiral arms.}
 \label{fig1}
\end{figure}

\begin{figure}
\includegraphics[width=84mm]{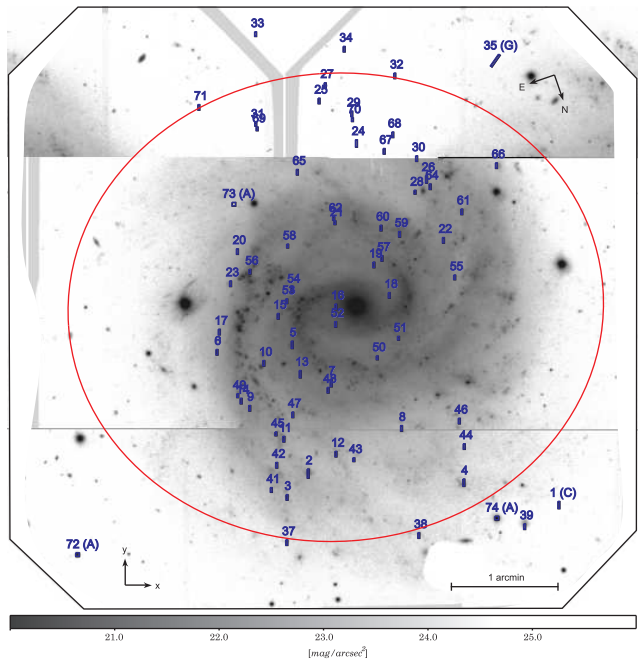} \caption{Slits and
elliptical fitting superimposed to the r\pr image of the galaxy
NGC1042. The same definitions used in the previous picture are used
here. Slits with a "G" sample galaxies in the field.}
 \label{fig2}
\end{figure}

\begin{figure}
\includegraphics[width=84mm]{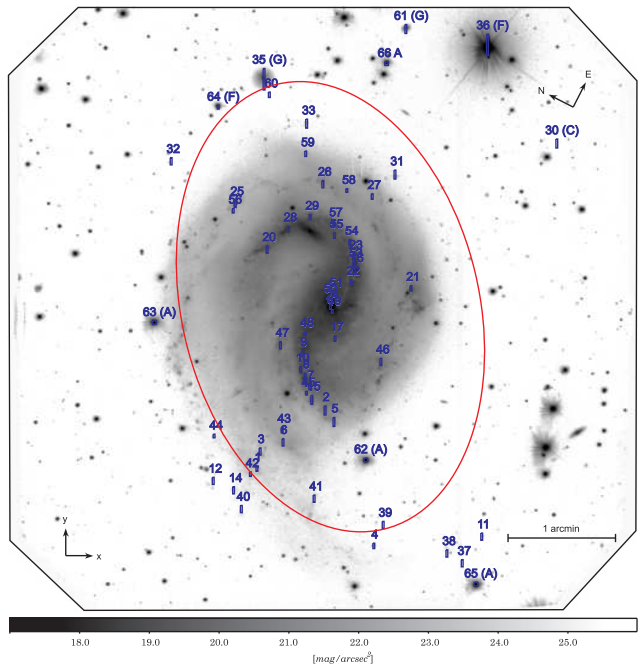} \caption{Slits and
elliptical fitting superimposed to the r\pr image of the galaxy
NGC6907. The definitions here follows those used in the two previous
pictures.}
 \label{fig3}
\end{figure}

 At first the pre-images (Figures \ref{fig1} to \ref{fig3}) were
observed in two filters:(g$^{\prime}$ and r$^{\prime}$). To
guarantee a continuous coverage of the GMOS field we applied
different dithering patterns to compose a single frame in each
filter. Since these images have better spatial resolution than the
public data and they are flux calibrated, we used them to refine the
\hii region selection, as described before. The combination of these
results with the morphological data allowed us to design the
metallic masks where the slits were placed for the spectroscopic
observation. Inclinations and position angles were estimated from
the isophotes of the final images of each galaxy. Table \ref{tbl-2}
summarizes the information for these observations.

\begin{table*}
 \caption{Observational parameters for the pre-imaging process of the
galaxies IC0167, NGC1042 and NGC6907 in each one of the filters g\pr
and r\pr. $UT$ is the universal time when the observation started,
$t_{exp}$ is the exposition time, $n_{exp}$ is the number of
exposures for the dithering pattern, $PA_{instr}$ is the
instrumental position angle and $OIWFS$ is the USNO UCAC2 Catalog
index of the star used as the on-instrument wavefront sensor. All
the observations were conducted at 2$\times$2 binned mode.}
 \label{tbl-2}

\begin{tabular}{lrrrrrr}
 \hline
  \hline
\multicolumn{ 1}{c}{{\bf }} & \multicolumn{ 2}{c}{{\bf IC0167}} & \multicolumn{ 2}{c}{{\bf NGC1042}} & \multicolumn{ 2}{c}{{\bf NGC6907}} \\
 \hline
  \hline
\multicolumn{ 1}{c}{{\bf Parameter}} &   \multicolumn{ 1}{c}{\bf g\pr} &   \multicolumn{ 1}{c}{\bf r\pr} &   \multicolumn{ 1}{c}{\bf g\pr} &   \multicolumn{ 1}{c}{\bf r\pr} &   \multicolumn{ 1}{c}{\bf g\pr} &   \multicolumn{ 1}{c}{\bf r\pr} \\
 \hline
\multicolumn{ 1}{l}{{\it UT}} &  16/8/2005 &  16/8/2005 &   6/9/2005 &   6/9/2005 &   1/8/2005 &   1/8/2005 \\
\multicolumn{ 1}{l}{{\it }}   &   14:31:36 &   14:38:57 &   12:58:18 &   13:09:54 &   08:03:12 &   08:09:04 \\
{\it t$_{exp}$ [s]}           &        150 &        120 &        240 &         90 &        360 &        120 \\
{\it n$_{exp}$}               &          5 &          4 &          8 &          3 &         12 &          4 \\
{\it PA$_{instr}$ [\dg]}      &        265 &        265 &         19 &         19 &        243 &        243 \\
{\it Airmass}                 &      1.001 &      1.001 &      1.179 &      1.163 &      1.691 &       1.66 \\
{\it Seeing [\dpr]}           &       0.99 &       0.74 &       0.78 &       0.68 &       0.63 &       0.63 \\
{\it OIWFS Star}              &   39427277 &   39427277 &   28845687 &   28845687 &   21894023 &   21894023 \\
 \hline
  \hline
\end{tabular}

\end{table*}

\par After the mask design the spectroscopic observations were carried
out doing individual exposures from 960 to 1200 seconds using the
B600$\_$G5303 grism centred at two different wavelengths to cover
the spectral range from 3500 {\AA} to 7100 {\AA}. With this coverage
and considering the systemic velocity of the galaxies the main
spectral lines required for metallicity evaluation could be
observed. No spectral dithering was used to cover the GMOS gaps in
the final spectra. So, taking this into account, the choice of \hii
regions only considered the slits in positions that would not
generate important spectral lines on the detector gaps.  A CuAr arc
lamp and a flat-field frame were observed in the sequence of the
spectroscopic acquisition, while the telescope was still following
the source, to guarantee the quality of the calibrations by avoiding
any effect caused by instrumental flexures. The same OIWFS stars
used in the pre-image process were used for the spectroscopy.
Information about the spectroscopic observations can be found in
Table \ref{tbl-3} and the details of the slits in Tables
 \ref{tblslit-1} to \ref{tblcorresp-2}.

\begin{table*}
 \caption{Observational details for the GMOS spectroscopy of the
galaxies IC0167, NGC1042 and NGC6907 in 2005 and 2006. $UT$ is the
universal time when the observation started, $t_{exp}$ is the
exposition time $\lambda_{c}$  is the central wavelength of the
dispersed light, $Band$ is the spectral range coverage, $n_{slit}$
is the total number of slits. The instrumental position angle was
the same used for the pre-images and an 1 arcsec slit width was
adopted in all observations. In 2005 the height of the slits were
fixed in 5 arcsec but in 2006 their sizes changed to improve the
number of observed \hii regions. The spectral resolution achieved is
around 1700.}
 \label{tbl-3}

\begin{tabular}{|l|rr|rr|rr|}
 \hline
  \hline
\multicolumn{ 1}{|c|}{{\bf }} & \multicolumn{ 2}{|c}{{\bf IC0167}} & \multicolumn{ 2}{|c}{{\bf NGC1042}} & \multicolumn{ 2}{|c}{{\bf NGC6907}} \\
 \hline
  \hline
\multicolumn{ 1}{|c|}{{\bf Parameter}} & \multicolumn{ 1}{|c|}{\bf 2005} & \multicolumn{ 1}{|c|}{\bf 2006} & \multicolumn{ 1}{|c|}{\bf 2005} & \multicolumn{ 1}{|c|}{\bf 2006} & \multicolumn{ 1}{|c|}{\bf 2005} & \multicolumn{ 1}{|c|}{\bf 2006} \\
 \hline
{\it UT}                  &   6/9/2005 &  23/8/2006 &  5/11/2005 &  24/8/2006 &   2/9/2005 &  22/8/2006 \\
{\it }                    &   11:40:27 &   14:41:54 &   10:34:08 &   14:56:00 &   06:04:33 &   09:30:27 \\
{\it t$_{exp}$ [s]}       &        960 &       1200 &       1200 &       1200 &        960 &       1200 \\
{\it $\lambda_{c}$ [\AA]} &       5000 &       5900 &       5000 &       5850 &       5000 &       5950 \\
{\it Band [\AA]}          &  3500-6300 &  4400-7200 &  3500-6300 &  4350-7150 &  3500-6300 &  4450-7250 \\
{\it n$_{slits}$}         &         34 &         52 &         35 &         68 &         36 &         57 \\
{\it Airmass}             &       1.06 &      1.018 &      1.154 &      1.137 &      1.618 &      1.453 \\
{\it Calib. Star}         &    G191B2B &    Feige34 &    G191B2B &    Feige34 &    G191B2B &    Feige34 \\
\hline
 \hline
\end{tabular}

\end{table*}

\section{Data Reduction and Analysis}

\par Standard procedures for reduction of GMOS imaging and spectroscopy
were followed. In both cases the general Gemini Data Reduction
Software was used. The details about this can be found in
\cite{Sca2008a} for the specific case of NGC6907. The
same procedures were carried out for IC0167 and NGC1042.

\par For each filter the multiple dithered frames were mosaiced
and co-added. This procedure improved the signal-to-noise ratio of
the final images removing both cosmic rays and the gaps between the
GMOS detectors. Since the spectroscopic data corresponds to a single
frame, the cosmic rays events were synthetically removed using the
task GSCRREJ (more details about in the task documentation). In
these cases the GMOS gaps were linearly interpolated. Wavelength
calibration was performed using the individual arc lamps observed
just after the spectroscopic acquisition. For each individual slit
the spatial distribution was used to improve the spectral
calibration. The sky lines and the continuum were removed from the
main source using the contiguous data in the spatial direction of
each slit.

\subsection{Photometry}

\par The photometric calibration in filters g\pr and r\pr was
performed using the photometric standard stars supplied by the
Gemini baseline, listed in Table \ref{tbl-3}. Usual procedures
applying DAOPHOT IRAF tasks enabled us to recover the zero point
magnitudes, taking into account the effects of the atmospheric
extinction and the corrections for the galactic extinction by
\cite*{Schlegel98} and \cite{Amores05}. Assuming the limit of
detection of the galaxies as the $3\sigma$ limit of dispersion of
the local sky level, elliptical fittings were performed for the
estimation of the inclination and position angle of the galaxies.
However, as next explained, we did not use the parameters obtained
in this way to derive the distances of \hii regions to the galactic
centres.

\par After the conversion of the g\pr and r\pr images to the B and V
bands, using the expressions by \cite{Fukugita96} and \cite{Hook04},
we obtained surface brightness profiles similar to those presented
by \cite{Canzian98}. However, the inclination and position angles
that we derive do not match with previous works like
\cite{Canzian98} and \cite{RC3}, even when we apply corrections such
as masking the data from the contribution of stellar fields, the
interarm regions and the disturbed parts of the galaxies. But this
is only part of reasons for giving preference to other methods to
determine inclinations and position angles. The method of fitting
ellipses to the isophotes, although commonly adopted, is based on
the questionable hypothesis that the spiral galaxies isophotes are
approximately circular when seen face-on. This is particularly
incorrect in the case of two-armed spirals with large pitch angles,
for which the light distribution is naturally elongated in the
direction of the extremities of the arms, the so-called Stock´s
effect \citep{Stock55}. The other methods available are the
kinematic method, which consists in finding the axes of the spider
diagrams or iso-velocity contours in \hi maps, and spiral-arm
fitting, which is based on the assumption that large parts of the
arms are logarithmic spirals. Our group performed \hi observations
with the GMRT (Giant Meterwave Radio Telescope) of the three
galaxies, but for the moment, the results are published only for
NGC6907 \citep{Sca2008a}, where a detailed discussion of its
inclination can be found. While the other results are not published,
the contradiction between photometric and kinematic results can be
seen from the orientation of the velocity fields presented for
NGC1042 by \cite{Kornreich00} and for IC0167 by \cite{vanMoorsel88}.
On the other hand, \cite{Ma01} proposed a method that uses the
spiral arms fitting, and applied it to several galaxies, including
IC0167 and NGC6907. Applying his method to both arms of NGC1042 we
got consistent values for inclination at a given position angle. We
found that the kinematic and spiral arm fitting methods give similar
results, which differ from the isophotal ellipse fitting. We
conclude that the spiral arm fitting method is reliable in the
present cases, and adopted its results in the following analysis
(Table \ref{tbl-4}).

\begin{table}
 \caption{Photometric and morphologic results for IC0167, NGC1042 and
NGC6907 using GMOS. The values with a "*" are from Ma (2001). For
the photometry, the symbol $m$ is used for total magnitude, $\Sigma$
for surface brightness, measured in mag/\sqas, $M$ is the absolute
magnitude and L the luminosity, in \lsun. For the morphology, $i$
and $\phi$ are the inclination and the position angles ($isoph$
index for isophotal elliptical fitting and $spir$ index for the
results derived from the spiral arms), $RA_{0}$ and $Dec_{0}$ are
the equatorial coordinates of the spiral arms centre (which coincide
with HI velocity center as mentioned in the text), $\alpha$ is the
pitch angle of the spiral arms, all in degrees and finally $a$ and
$b$ are major and minor semi-axis of the ellipse, in arcmin.}
 \label{tbl-4}
 \centering
\begin{tabular}{lrrr}
 \hline
{\bf Parameter}          & {\bf IC0167}        & {\bf NGC1042}       & {\bf NGC6907}       \\
 \hline
{\it $m_{0,g}$}        & 31.49 $ \pm $ 0,15  & 31.47 $ \pm $ 0,16  & 31.4 $ \pm $ 0,15   \\
{\it $m_{0,r}$}        & 31.79 $ \pm $ 0,17  & 31.77 $ \pm $ 0,21  & 31.72 $ \pm $ 0,14  \\
{\it $m_{g}$}            & 11.11 $ \pm $ 0,29  & 9.84 $ \pm $ 0,20   & 10.45 $ \pm $ 0,04  \\
{\it $m_{r}$}            & 10.26 $ \pm $ 0,27  & 9.16 $ \pm $ 0,19   & 8.90 $ \pm $ 0,03   \\
{\it $\Sigma_{g}$]}      & 21.95 $ \pm $ 0,57  & 21.80 $ \pm $ 0,44  & 21.3 $ \pm $ 0,08   \\
{\it $\Sigma_{r}$}       & 21.17 $ \pm $ 0,56  & 20.95 $ \pm $ 0,44  & 19.7 $ \pm $ 0,07   \\
{\it \MB}                & -21.07 $ \pm $ 0.47 & -20.63 $ \pm $ 0.41 & -21.70 $ \pm $ 0.35 \\
{\it \LB [10$^{10}$}     & 3.9 $ \pm $ 1.7     & 2.6 $ \pm $ 1.0     & 7.0 $ \pm $ 2.2     \\
{\it $\Sigma_{lim}^{g}$} & 26.26 $ \pm $ 0,57  & 26.33 $ \pm $ 0,44  & 25.74 $ \pm $ 0,08  \\
{\it $\Sigma_{lim}^{r}$} & 25.44 $ \pm $ 0,56  & 25.4 $ \pm $ 0,44   & 24.86 $ \pm $ 0,07  \\
{\it $RA_{0}$}           & 27.785629           & 40.099802           & 306.277600          \\
{\it $Dec_{0}$}          & 21.912776           & -8.433441           & -24.809200          \\
{\it $a_{isoph}$}                & 1.82 $ \pm $ 0.04   & 2.49 $ \pm $ 0.15   & 1.47 $ \pm $ 0.02   \\
{\it $b_{isoph}$}                & 1.15 $ \pm $ 0.03   & 2.01 $ \pm $ 0.10   & 1.29 $ \pm $ 0.02   \\
{\it $i_{isoph}$}                & 51.5 $ \pm $ 3.0    & 36.3 $ \pm $ 3.1    & 29.1 $ \pm $ 3.0    \\
{\it $\phi_{isoph}$}             & 73.1 $ \pm $ 15.8   & 12.5 $ \pm $ 8.5    & 254.1 $ \pm $ 3.3   \\
{\it $i_{spir}$}         & 38.0 $ \pm $ 3.0*   & 29.0 $ \pm $ 2.9     & 49.5 $ \pm $ 3.0*  \\
{\it $\phi_{spir}$}      & 134.0 $ \pm $ 3.0*  & 277.4 $ \pm $ 5.1    & 239.9 $ \pm $ 3.0* \\
{\it $\alpha$}           & 26.6 $ \pm $ 1.9*   & 13.3 $ \pm $ 2.5     & 23.5 $ \pm $ 5.0*  \\
 \hline
\end{tabular}

\end{table}

\par
\begin{figure*}
\includegraphics{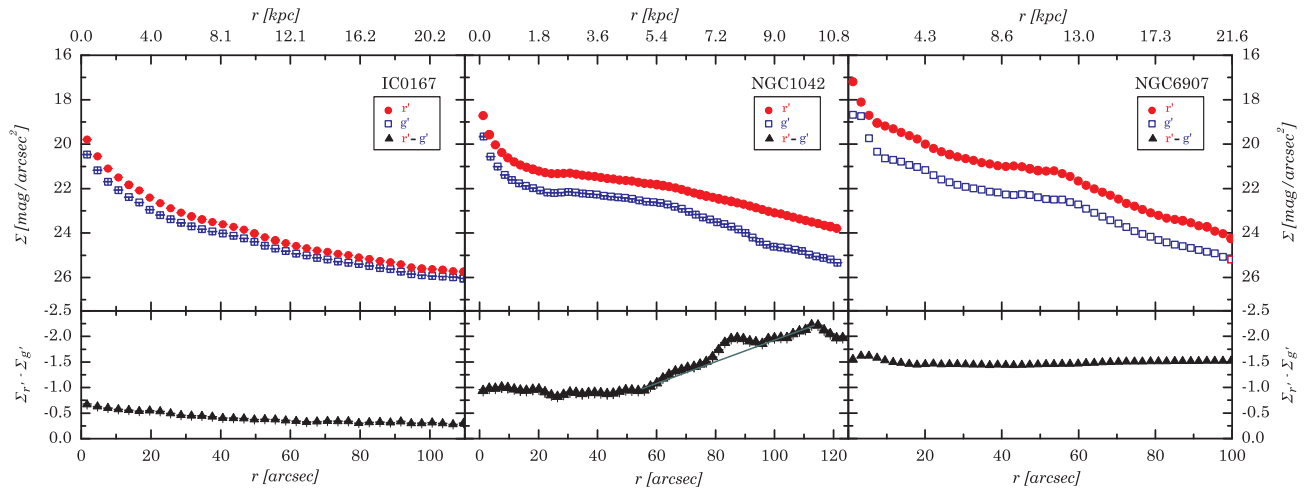} \caption{g$^{\prime}$ and
r$^{\prime}$ radial surface brightness (top) and color profiles
(bottom). Remark that in both the order of the filters and the scale
are not conventional.}
 \label{fig4}
\end{figure*}

 The radial surface brightness profiles (Figure \ref{fig4}) were
obtained performing the integration of the flux on elliptical rings
with the same thickness of the seeing and following the parameters
of projection imposed by the spiral arms (Table \ref{tbl-4}). Since
there is an interaction between NGC6907 and NGC6908, confirmed by
\cite{Sca2008a}, a detailed discussion of the photometry of NGC6907
can be found in that paper.

\subsection{Spectroscopy}

\par With the spectrophotometric calibration stars supplied by the
Gemini baseline (Table \ref{tbl-3}) we determined the sensitivity
function and then each spectrum could be flux calibrated.

\par Considering that all spectral lines have their wavelength
systematically changed by the systemic velocity of the galaxies, and
locally changed by the projection of  the rotation curves on the
line of sight, we used the IRAF package RVSAO \citep{Kurtz98} to
determine radial velocities and identify the typical spectral lines
observed in H~{\sc ii} regions by cross-correlating a template
spectrum to the observed spectra. The  measured velocities and their
errors are given in Tables \ref{tblslit-1} to \ref{tblslit-6}. We
then integrated the flux of lines, performing the following steps:

\begin{enumerate}
 \renewcommand{\theenumi}{(\arabic{enumi})}

\item According to the galactic coordinates of each slit the
related spectrum was corrected by the galactic extinction using the
reddening calculated by the code EXTIN written by \cite{Amores05}
and an IDL adaptation of the code FMCURVE by \cite{Fitzpatrick99}
called FM\_UNRED;

\item Using the expected ratios between Balmer lines
\citep{Osterbrock89} we measured the flux of the hydrogen lines and
estimated the intrinsic reddenings of each spectrum employing the
relations written by \cite{Fitzpatrick99}, which were introduced in
the FM\_UNRED code to correct the spectra by the intrinsic
reddening;

\item Taking into account the wavelengths determined for all recognized
spectral lines we performed the integration of the flux of them, by
considering the robust mean of the area under the line profile
evaluated by the following procedures: (a) Gaussian fitting; (b)
Lorentzian fitting; (c) numerical integration; (d) double gaussian
fitting;

\item With the fluxes and uncertainties evaluated in the previous
step, the metallicities were calculated using statistical methods to
determine {\it O} abundances. The methods applied were the R23
method of \cite{Pagel79} with the considerations of
\cite{McGaugh94}, the [O~{\sc iii}]/[N~{\sc ii}] method of
\cite{PP04} and \cite{Stasinska06}, the [N~{\sc ii}]/$H\alpha$
method of \cite{PP04}, the [Ar~{\sc iii}]/[O~{\sc iii}] method of
\cite{Stasinska06} and the {\it p-method} of \cite{Pilyugin00} and
\cite{Pilyugin01}.

\end{enumerate}

\par
\begin{figure*}
\includegraphics{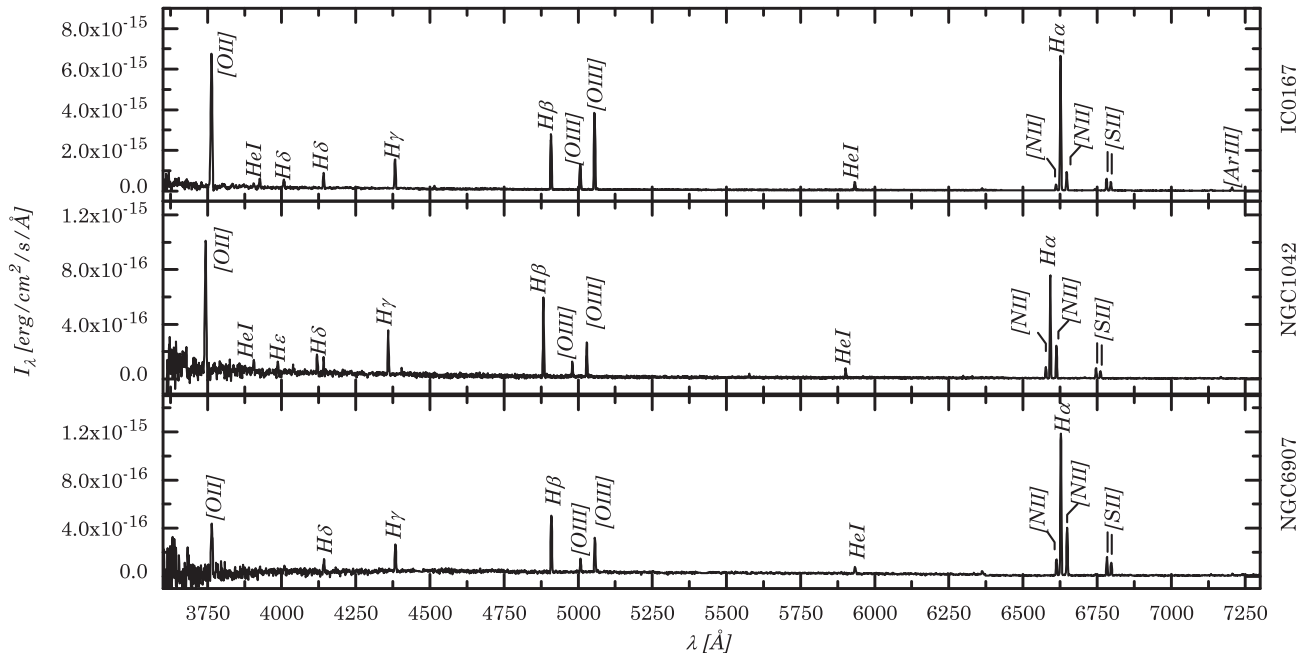} \caption{Typical spectra of the
galaxies IC0167 (top), associated to slit 18 (2006), NGC1042
(middle) from slit 21 (2006) and NGC6907 (bottom) related to the
slit 29 (2006). Each spectrum combines the data from 2005 and 2006
observations}
 \label{fig5}
\end{figure*}

 We next present more details on the steps described above.
Since the interstellar medium affects differentially the flux of the
spectral lines, the correction of the spectra by the extinction is
fundamental to estimate metallicities. To do this we need to
distinguish two contributions: one from our Galaxy (the galactic
extinction) and the other internal to the observed galaxies (the
intrinsic extinction). A model for the galactic extinction is well
established by \cite{Amores05} and we applied their model to correct
all spectra by the galactic flux losses. Supposing that the
remaining differences in the ratio of the Balmer lines are a
consequence of the intrinsic extinction in the observed galaxies and
that their extinction laws are similar to the Galaxy extinction law,
each spectrum was corrected again by determining the reddening from
the Balmer line ratios and using it in the \cite{Fitzpatrick99}
code. In spite of the evidences from the Megellanic Clouds that
extragalactic reddening laws may not be rigorously the same one we
observe for our Galaxy, the comparison between the results of
\cite{Gordon03} for the Megellanic Clouds with the
\cite{Fitzpatrick99} extinction law reveals that at least in the
optical part of the spectrum, where the {\it O} abundances are
calculated, this hypothesis may be assumed. This was already done in
the past (eg. \citealt*{Misselt99}, working with 11 spiral galaxies)
and it has the advantage of correcting the rising of the extinction
as a function of the galactic inclination \citep{Unterborn08}. The
Tables \ref{tblext-1} to \ref{tblext-6} contain the values we got
for intrinsic extinctions. Typical spectra of each one of the
observed galaxies corrected for both extinctions can be seen in
Figure \ref{fig5}. Ultimately the results would be the same if we
ignored the extinction components and only the total reddening
correction using the Balmer lines were considered.

\par The option of measuring the fluxes using four procedures was a way
to automatically register the differences between the spectral lines
profiles in each spectrum. The robust mean of these fluxes is
equivalent to the mean of the two best fitting procedures, since in
this method the "discrepant" values are discarded. In our case the
best fitting procedures were the Gaussian fits and the numerical
integration. But the mean between these two procedures is exactly
the method adopted by \cite{Zaritsky94} to measure the line fluxes
of his sample. The advantage of our method is that it allowed us to
automatically compare and identify profiles affected by cosmic rays
remnants, bad sky line subtraction, lines with unusual profiles and
blended lines. We adopted uncertainties associated to the robust
standard deviation of the measured fluxes in such way that larger
uncertainties are a consequence of less defined profiles and lines
with lower signal-to-noise ratio. These results were verified by
computing the Poissonian errors directly from the number of photon
counts in the lines and the results are in perfect agreement with
our method. Typically these values are smaller than 10\% but
considering differential effects each spectral line has its own
typical value (see Tables \ref{tblflx-1} to \ref{tblflx-3}).

One of these effects is the differential atmospheric refraction.
Depending on the airmass, seeing, slit width and slit orientation
relative to the parallactic angle the observations may be more or
less affected by this kind of refraction. According to our
specifications, observations were conducted in seeing conditions
better than 0.7", while the standard slit width of 1" was adopted
for all objects. Since for the MOS observations the position angles
of the slits are fixed and our observations were carried out in
queue mode, we were not able to tune the slits position angles to
the parallactic angles. Nevertheless, except for the observation of
NGC6907 in 2005 all observations were performed with position angles
of the slit differing less the 15 degrees from the parallactic
angle. Unfortunately NGC6907 is always observed in large airmasses
at Mauna Kea, and this affected both observations of this galaxy.

Considering only the differential wavelength effect of the airmass
on the atmospheric refraction, we verified that the [O~{\sc ii}]
spectral line was severely affected in the observations of NGC6907
in 2005. For these observations the center of the \hii regions were
deviated at least 0.5" outside the slit width; this means that the
observed [O~{\sc ii}] lines come predominantly from the borders of
the \hii regions. Moreover, the slit orientation relative to the
parallactic angle in 2005 was almost 80 degrees, which means that
more than 85\% of the [O~{\sc ii}] emission line was not detected.

Fortunately the situation was different for all the other
observations. Taking into account the redshifts of the galaxies, the
worst airmasses and the largest deviation of the slit orientation
relative to the parallactic angle, the [O~{\sc ii}] flux losses
represents less than 5\% of the total [O~{\sc ii}] emission. This is
the particular situation of NGC1042 observed in 2005. Since IC0167
was observed almost in the parallactic angle and airmass 1, there
was no effect of atmospheric refraction. Furthermore, for the 2006
observations we only used spectral lines above $H\gamma$, for which
the atmospheric refraction effects are negligible given our slit
widths.

Another thing that must be considered is the underlying Balmer
absorption. The corrections are usually done by measuring the
equivalent widths of the Balmer lines and applying the procedures by
\cite*{McCall85} and \cite{Oey93}. According to \cite{Zaritsky94}
such correction amounts to less than 10\% for 75\% of the regions he
studied. Given that for most of the spectra we observed there is no
evident absorption profiles in $H\alpha$, $H\beta$ or $H\gamma$
(used to correct the spectra for reddening and as references for
abundance determinations) we opted to eliminate from our analysis
the objects for which the $1\sigma$ level of the background affected
by the Balmer absorption represents more than 10\% of the peak
emission. It represents less than 20\% of the observed regions in
2005 and about of 10\% of our sample observed in 2006. Given the
proportions presented by Zaritsky we believe that the worst effect
of the Balmer absorption was eliminated from our sample.

\section{Results and Discussion}

\par The first step to determine {\it O} gradients is to evaluate the
{\it O} abundances and the radial distances inside de galactic
planes. Following the procedures mentioned in the previous section,
the flux of each spectral line needed to calculate the {\it O}
abundances using the statistical methods was determined and
registered in Tables \ref{tblflx-1} to \ref{tblflx-3}. Some
individual lines may present underestimated uncertainties,
nevertheless the errors on the abundances, derived by propagating
the flux uncertainties of all spectral lines involved, are around
0.2 dex, coherent with those mentioned in the literature (see
\citealt*{Pilyugin2004} and references therein). Using the
inclinations, position angles and the galactic centre coordinates
determined using the spiral arms (Table \ref{tbl-4}), the distances
to the \hii regions sampled by the slits (Tables \ref{tblslit-1} to
\ref{tblslit-6}) were calculated with the expressions given by
\cite{Sca2008a}. These expressions introduce important corrections
to the more traditional ones (eg. \citealt{Begeman87}). For the
galaxies studied here these corrections reach up 20\% of the
distances calculated with the previous procedures. Since the
deviations on the corotation radius range between 2 and 3 kpc the
corrections are needed to avoid a too large smoothing of the radial
distribution of metallicity.

\par For each galaxy the systemic velocities were obtained by taking
into account the radial velocities measured from all the observed
spectra and calculating the weighted mean of their distribution.
Such procedure is a raw measurement, which assumes that the receding
and the approaching sides of the galaxies were evenly sampled. In
spite of that, the results are in agreement with those found in the
literature (Table \ref{tbl-1}). A few objects were found to be
background galaxies or isolated objects instead of \hii regions
belonging to the galaxies we study here; they were eliminated from
the sample based on their velocities. The velocity distribution of
the \hii regions of each galaxy is approximately Gaussian, and we
discarded the objects more than 3 HWHM (half width at half maximum)
off from the center of the distribution.

\par
\begin{figure*}
\includegraphics{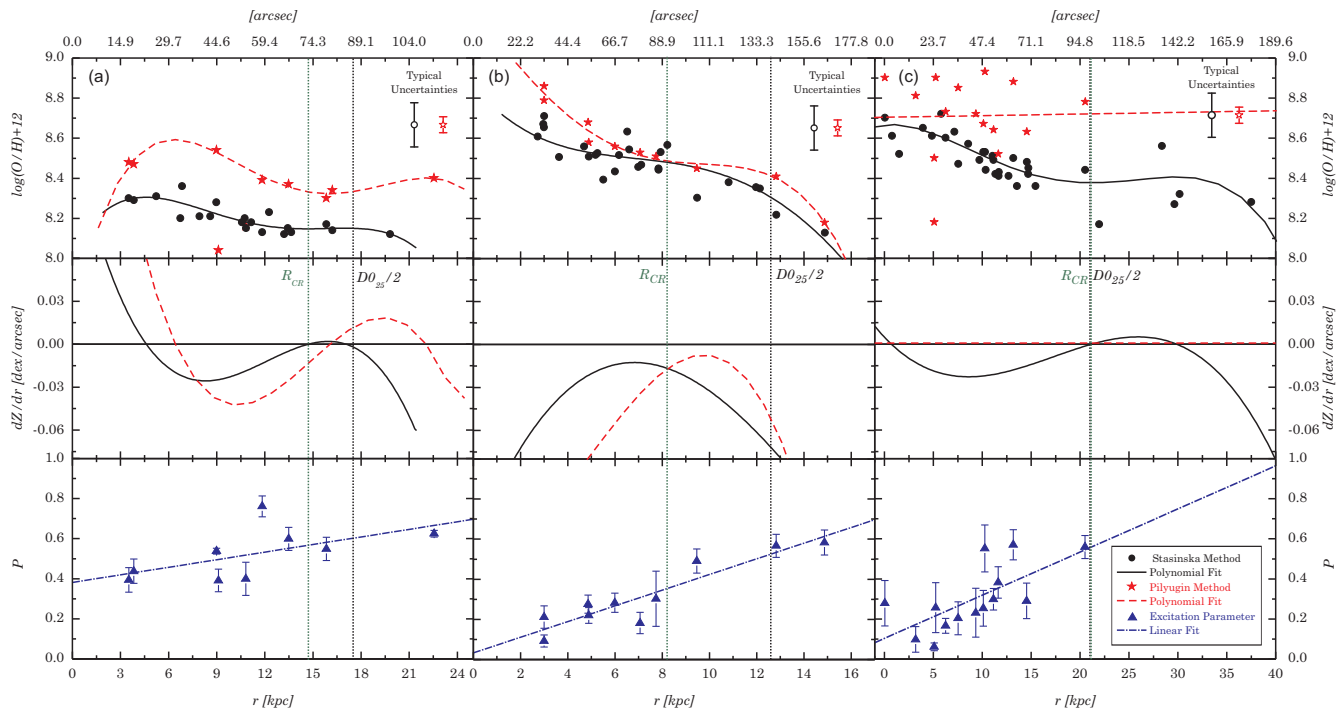} \caption{ Radial {\it O} abundance, fitted by
a 4th order polynomial (top), the derivative of the fitted curve
(middle) and the distribution of the excitation parameter {\it P}
(bottom). The stars represent the results of the 2005 observations,
analyzed with the Pilyugin's method, and the circles symbolize the
observations of 2006, analyzed using the Stasi{\'n}ska's method. For
each case the opened symbols indicated the typical uncertainties.
The vertical lines labeled with a $R_{CR}$ indicated the estimated
position of the corotation radius and those labeled with a
$D0_{25}/2$ mark the position of the isophotal radius where the
surface brightness falls to $25 mag/arcsec^{2}$. The panels are: (a)
IC0167 (left), (b) NGC1042 (center), (c) NGC6907 (right). In this
last case the differential refraction leads to a so huge scatter
when the Pilyugin technique was applied, so it was not used in the
further interpretations.}
 \label{fig6}
\end{figure*}

 Remembering that each observational run was optimized for a
given spectral range we present in Figure \ref{fig6} the best {\it
O} abundances calculated for each galaxy using the spectral lines
available in that run. Since in 2005 observations the [O~{\sc ii}]
line was present and $H\alpha$ was not, we applied the R23
based-methods to determine {\it O} abundances, represented here by
the {\it p-method} (\citealt{Pilyugin00} and \citealt{Pilyugin01}).
On the other hand, for the 2006 observations, the [O~{\sc ii}] line
was out of the range and only the methods that use the $H\alpha$ and
$H\beta$ lines could be applied, in particular [O~{\sc iii}]/[N~{\sc
ii}] methods, represented here by the \cite{Stasinska06} method. The
advantage of the {\it p-method} is that it can provide {\it O}
abundances with precision comparable with the direct methods, based
on the temperature sensitive line ratios, however it is
double-valued at low abundances. Note that the 2005 observations
could only be corrected for extinction by measuring the $H\beta$ to
$H\gamma$ ratio, since $H\alpha$ was not available. This provides a
poorer correction than the  $H\alpha$ to $H\beta$ ratio.  In
contrast to the {\it p-method}, the [O~{\sc iii}]/[N~{\sc ii}]
Stasi{\'n}ska's method is single valued, with the advantage that it
is relatively independent of the reddening \citep{Stasinska06}.

\par A constant gradient has been the most commonly adopted law in
the study of metallicity of galaxies. This is a first approximation,
not a choice dictated by theory. For instance, most chemical
abundance models adopt star formation rates proportional to some
power of the gas density, and most often the gas density presents a
peak at several kpc from the center.  The stellar density in the
disk of our Galaxy is better described by Kormendy's function than
by an exponential law (see \citealt{Lepine00}). We make here the
choice of fitting the data with 4th order polynomials. This does not
correspond to a theoretical model of galactic structure; the
polynomial fitting is only adopted as a smoothing method instead
other methods like splines or Gaussian filtering, which could be
used alternatively. One advantage of the polynomial method is that
it gives the position of minima and inflections in an easy way. The
choice of the 4th order corresponds to the amount of details that we
are willing to reveal. For instance in our Galaxy, the observations
of Cepheids (\citealt{Andrievsky04}) show that there is a strong
metallicity gradient in the inner regions, followed by a plateau,
followed again by a another strong gradient. Such a behavior, that
we expect to observe in other galaxies as well, can be approximated
by a 4th order polynomial.

\par It is known that the different statistical methods used
to calculate the {\it O} abundances can result in different values
for a same observation, and they can produce different dispersion of
the data with respect to an average abundance profile, and also
small differences in the gradients, specially for those methods
which do not depend directly on the {\it O} ionic abundances. In our
observations the largest differences occurred between the {\it
p-method} and all the other mentioned methods. This seems to be a
consequence of the strong dependence of the {\it p-method} on the
[O~{\sc ii}] emission line, which is observed at the limit of the
instrumental sensitivity and it is strongly affected by the
reddening and the parallactic angle misalignments.

\par For IC0167 most of the methods applied here point towards the
existence of a very smooth minimum in the {\it O} abundance profile,
which is shallower than the one of our Galaxy, according to
\cite*{Mishurov02}. In particular, the distribution produced by the
{\it p-method} is more dispersed than the results for the [O~{\sc
iii}]/[N~{\sc ii}] method, however  the minima in the {\it O}
abundance obtained with the two methods almost coincide. The median
of the minima and inflections found using all the methods applied to
this galaxy is $14.7 \pm 2.6$ kpc. Obviously it is necessary to be
cautious in the interpretation of the fitted curves. The trend could
be fitted by a straight line and the evidence for an inflection is
not very strong in this case.

\par On the other hand, for NGC1042, the abundance dispersion is less
pronounced and a plateau is present in the {\it O} abundance
profiles of this galaxy obtained by means of the [O~{\sc
iii}]/[N~{\sc ii}] {\it p} and R23-based methods. This plateau could
possibly be interpreted as being a consequence of the corotation as
well, but in this case the minimum is too shallow to contrast with
the dispersion of the data. The dispersion resulting from the two
calibrations of [N~{\sc ii}]/$H\alpha$ method of \cite{PP04} and the
[Ar~{\sc iii}]/[O~{\sc iii}] is so high that no consistent
information can be extracted. Since the point of inflection is a
point which is close to being a minimum (see the lower curves in
Figure \ref{fig6} for NGC1042), we could speculate that it is also
an indicator of the corotation. The median of the inflexions of the
abundance curves for NGC1042 is $8.2 \pm 2.8$ kpc. Interestingly, at
this same position a small bump also can be observed in the color
distribution of this galaxy (Figure \ref{fig4}). This might be
connected with  corotation as well, since a gap in the density
distribution of young stars can be expected at corotation, as it
happens in our Galaxy \citep*{Amores09}. The lack of young stars
would  make this region slightly redder than its neighborhood.

\par Finally, a bimodal behavior in the radial distribution of
{\it O} is found for NGC6907 using the [O~{\sc iii}]/[N~{\sc ii}]
method. Since the Argonium line is not detected along all the
galaxy, the spatial distribution of metallicity derived from this
element is inconclusive. The same happens using the [N~{\sc
ii}]/$H\alpha$ method, which results in {\it O} abundances with
large dispersion at radii beyond 15 kpc. Unfortunately the
observations of 2005 were not performed near the parallactic angle
and for this reason the measurements of the [O~{\sc ii}] lines are
not reliable, causing a huge dispersion of the {\it O} abundances
derived by the R23 and {\it p-method}. Since the nucleosynthetic
origin of the nitrogen may shift from secondary to primary in low
abundances \hii regions we need to be more cautious to interpret the
inflexions in the NGC6907 abundance distribution. The median of the
radii of the minima found using the other statistical methods is
$21.1 \pm 3.7$ kpc. \cite{Sca2008a} have shown that for this galaxy
there is a minimum in the \hi gas distribution between 20 and 30
kpc, which can be related with the same effect of the corotation
found in our galaxy by \cite{Amores09}.

\par The interpretation of the gradient of metallicity for NGC6907
may seem dangerous, since this galaxy is interacting with NGC6908 as
discussed by \cite{Sca2008a}. However the same authors have shown
that the influence of this galaxy on the NGC6907 gas distribution is
restricted to about 20\dg around NGC6908, in azimuthal angles
measured in NGC6907 galactic plane. Except for the spectra sampled
by the slit 25 in the 2006 run, all the other slits associated to
radii greater than 15 kpc are situated on the side opposite to the
one where the interaction with NGC6908 takes place. Similarly to
what happens with the rotation curve, which is not perturbed on that
side \citep{Sca2008a}, we expect that the metallicities of the \hii
regions presented here are not affected by the interaction.

\par Are the polynomial fits really better than the traditional straight
line fits? It should be remembered that we do not consider that the
4th order polynomials are models, but only a technique used for
locating changes in the slope. If the changes in slopes and plateaux
are real then, in principle, the polynomial fit should be better.
But of course, a 4th order polynomial always produces a better fit
than straight line, as long as the rms deviations of the data with
respect to the fitted curve are taken as the measure of the quality
of the fit. To determine if a polynomial fit is effectively a
meaningful choice, one must compare the chi-square divided by the
degree of freedom $N-k-1$, where $N$ is the number of data points
and $k$ the order of the polynomial (see the discussion in Chapter
15 of Numerical Recipes (\citep{numrec1992}). We performed these
tests, and the results can be summarized as follows: after
correction for the degree of freedom, the adjustments by 4th order
polynomials are slightly better (the chi-square obtained with
straight lines are larger by a factor which ranges from 2 to only a
few percent). In other words the straight lines are not the best
choice, but they cannot be excluded on statistical grounds. Testing
different orders of polynomials from 2 to 6, in all cases the 4th
order is slightly better than all the others except for NGC1042 for
which the 3rd order polynomial gives a slightly smaller (3\%)
normalized chi-square. This last result still suggests that an
inflection point is present.

\cite{Pilyugin03} argued that the bends in the slopes of radial
abundance gradients in the disks of spiral galaxies is a consequence
of a systematic error involving the excitation parameter P, which
would produce realistic {\it O} abundance in high-excitation \hii
regions, but would  overestimate this abundance in low excitation
\hii regions. For a measured value of R23, there are two possible
solutions for the {\it O} abundance. For instance, according to
Pilyugin, the positive slope of the observed {\it O} abundance by
\cite{RoyWalsh97} in the external regions of the galaxy NGC1365
would be changed into a negative slope, and the break would
disappear, if one used the lower branch of the graph of {\it O}
abundance versus R23 for these regions. The question, then, is when
one should move from one branch to the other, to use correctly the
method. According to Pilyugin's calibrations \citep{Pilyugin03}, the
upper branch corresponds to $12+log(O/H) > 8.15$. As can be seen in
Figure \ref{fig6}, the slope breaks  are situated above this
abundance threshold for two of the galaxies studied here (NGC1042
and NGC6907), so they cannot be explained by a change of the {\it O}
abundance versus R23 law. One argument given by Pilyugin to justify
the existence of bends above the previous limit is that the less the
value of the excitation parameter {\it P} the more the oxygen
abundance obtained with the R23-method is overestimated
\citep{Pilyugin03}. However, even with an increasing behavior of the
excitation parameter at radii larger than the radius identified as
the corotation radius for NGC1042, the overall decreasing behavior
of the radial metallicity distribution is not flattened by the
almost linear radial increase of the excitation parameter {\it P}.
It only presents a small step in the region where the corotation
seems to occur. Another important point to be considered is that, in
contrast to the R23 or the {\it p-method} the [O~{\sc iii}]/[N~{\sc
ii}] method is single-valued and for the galaxies observed here we
verified minima and inflexions independently of the R23-based
methods. One might worry that [O~{\sc iii}]/[N~{\sc ii}] could
depend upon a second parameter near co-rotation, and the behavior of
the excitation parameter might even suggest a correlation with the
corotation radius, but such an hypothesis would have to be
investigated in a large sample of galaxies, and for the moment, it
is not supported by any theoretical prediction.

This is not the first time that breaks in the radial distribution of
metallicity are associated to the corotation radius. The break
observed by Roy \& Walsh in NGC1365 is real, as it has been observed
using different calibrations and not only R23, and in addition, the
break appears at $12 + log(O/H) = 8.7$, well in the upper branch
defined by Pilyugin. \cite{RoyWalsh97} verified that the position of
the metallicity break that they found in NGC1365 is beyond the
corotation radius found by \cite{Jorsater95}. Remembering that the
Mishurov's model for our Galaxy, as well as the more recent one by
\cite{Acharova10}, claims that the chemical evolution of a galaxy is
proportional not only to the difference between the spiral pattern
speed and the angular velocity of the rotation curve, but also
depends on the surface density of the gas, so a small displacement
of the metallicity break with respect to the corotation radius can
be understood. In NGC1365, there is an intense star formation
activity in the bar, at a radius slightly smaller than the
corotation radius, in a region where the gas density is higher than
average. Slightly beyond the corotation radius, there is steep
decline in the gas density \citep{Jorsater95}. The presence of a
strong gas density gradient across the corotation radius displaces
the metallicity minimum simply because the star-formation rate
depends strongly on that density (see the details of Mishurov's
model).

\section{Conclusions}

The analysis of the gradient of metallicity of our Galaxy by
\cite{Mishurov02} leads us to infer that in other galaxies as well,
we should expect that  minima or inflexions in the radial gradient
of metallicity could be correlated with the corotation radius. One
indication that the same process also happens to other galaxies was
obtained with the correlation found between the minima and
inflexions in the metallicity distribution, using published data and
the procedures presented here, and the corotation radius from
different references available from the literature \citep{Sca2010}.
In order to test this prediction we performed photometric and
spectroscopic observations of three of the best candidates pointed
by \cite{Canzian98} to have the corotation inside the optical disk,
IC0167, NGC1042 and NGC6907.

Using images obtained with the Gemini Telescope, we conducted a
careful procedure to select \hii regions to be observed with the
GMOS. To avoid the problems with the isophotal elliptical fittings
found for these galaxies, caused by their prominent spiral
structure, we adopted the procedure recommended by \cite{Ma01} to
determine the inclination and the position angles. This allowed us
to calculate more reliable distances in the galactic plane, since
for these galaxies the spiral arms provide results which are
coherent with the velocity fields found by \cite{Sca2008a},
\cite{Kornreich00} and \cite{vanMoorsel88}.

The first spectroscopic run was optimized for the blue part of the
spectra, for which the R23-based methods could be applied, in
particular the {\it p-method}, while the second run was optimized to
the red part of the spectra, and the [N~{\sc ii}]-based methods
could be employed. The number of observed and effectively used
spectra in this project changed according to the run but with the
observations of 2006 we overcome the minimum data sample advocated
by \cite{DutilRoy01} for an ideal measurement of the metallicity
gradient. All data were corrected for galactic and intrinsic
extinctions and the data affected by the underlying Balmer
absorption was removed from the analysis. By fitting polynomials of
different orders to the radial metallicity distribution of the three
galaxies, we found that the 4th order polynomials produces better
fits than first order (straight lines) in all the cases, even when
correcting the chi-square for the larger number of degrees of
freedom of the 4th order polynomial. The radii of the minima or
points of inflection derived from the statistical methods of
determining the abundances adopted here are in relatively good
agreement. We recognize, however, that since the gradients and their
variations have small amplitudes, we cannot totally exclude the
possibility of an interpretation of the data in terms of straight
lines. Furthermore, since the Canzian´s method only gives us the
best candidates of galaxies that have the corotation radii inside
the optical disk, it is not possible for the moment to clearly
associate the changes in gradient slopes with corotation. If we
assume that the minima and inflections are situated at the
corotation radius, for IC0167, NGC1042 and NGC6907, the corotation
is at $14.7 \pm 2.6$, $8.2 \pm 2.8$ and $21.1 \pm 3.7$ kpc
respectively. Further studies have to be conducted to verify if
these estimates of corotation radius are confirmed by other methods.

\section{Acknowledgements}
\par This work is based on observations obtained at the Gemini Observatory, which
is operated by the Association of Universities for Research in
Astronomy, Inc., under a cooperative agreement with the NSF on
behalf of the Gemini partnership: the National Science Foundation
(United States), the Science and Technology Facilities Council
(United Kingdom), the National Research Council (Canada), CONICYT
(Chile), the Australian Research Council (Australia), Minist\'erio
da Ci\^encia e Tecnologia (Brazil) and Ministerio de Ciencia,
Tecnología e Innovaci\'{\o}n Productiva  (Argentina) The work was
supported by the Sao Paulo State Agency FAPESP through the grant
09/05181-8. This research has been benefited by the NASA's
Astrophysics Data System (ADS) and Extra-galactic Database (NED)
services. Their open software used in this research is greatly
acknowledged. The Gemini programme ID for the data used in this
paper is the GN-2005B-Q-39 and GN-2006B-Q-89 (PI: S. Scarano Jr). We
would like to thank the referee Marshall McCall for his careful and
critical reading of the earlier versions of this paper and for his
numerous valuable and substantial suggestions for improvements of
the paper.

\bibliographystyle{mn2e}
\bibliography{gradmetal}

\clearpage
\appendix
\section{Positions, Distances and Velocities Associated to the Slits}

\par In this appendix we compile the information about the sky
coordinates of the slits and its height (RA (J2000), Dec (J2000),
$\Delta$y), the correspondent face-on polar coordinates measured
from the kinematic centre of the galaxies (r, $\theta$), the
cross-correlated velocities evaluated from the observed spectra
(v$_{obs}$), the velocity dispersion ($\sigma_{v_{obs}}$) and the
number of lines detected (n$_{lin}$) and effectively used on the
calculations (n$_{us}$).

\begin{table*}
 \caption{Information about the spectra observed for IC0167 in 2005.}
 \label{tblslit-1}
  \centering
\begin{tabular}{ccccrrrrcc}
\hline \hline
\multicolumn{ 10}{c}{{\bf IC0167 - Generic Spectral Information on the Slit - 2005 Observation}} \\
\hline \hline
\multicolumn{ 1}{|c|}{{\bf ID}} & {\bf RA}    & {\bf Dec}   & {\bf $\Delta$y}                 & \multicolumn{ 1}{c}{\bf r}      & \multicolumn{ 1}{c}{\bf $\theta$} & \multicolumn{ 1}{c}{\bf v$_{obs}$} & \multicolumn{ 1}{c}{\bf $\sigma_{v_{obs}}$} & {\bf n$_{us}$} & {\bf n$_{lin}$} \\
\multicolumn{ 1}{|c|}{}         & {\bf [\dg]} & {\bf [\dg]} & \multicolumn{ 1}{c}{\bf [\dpr]} & \multicolumn{ 1}{c}{\bf [\dpr]} & \multicolumn{ 1}{c}{\bf [\dg]}    & \multicolumn{ 1}{c}{\bf [km/s]}    & \multicolumn{ 1}{c}{\bf [km/s]}             &                &                 \\
\hline
   {\bf 1} &   27.81954 &   21.94981 &          5 &        222 &        -88 &       1620 &        160 &          5 &          8 \\
   {\bf 2} &   27.82779 &   21.92064 &          5 &        171 &        -52 &      74130 &         46 &          9 &         11 \\
   {\bf 4} &   27.81574 &   21.94136 &          5 &        183 &        -84 &          - &          - &          - &          - \\
   {\bf 5} &   27.80700 &   21.92086 &          5 &         95 &        -62 &       2962 &         47 &         12 &         14 \\
   {\bf 6} &   27.80872 &   21.92206 &          5 &        104 &        -63 &       2937 &         19 &         11 &         13 \\
\hline \hline
\end{tabular}
\end{table*}

\begin{table*}
 \caption{Information about the spectra observed for IC0167 in 2006.}
 \label{tblslit-2}
  \centering
\begin{tabular}{ccccrrrrcc}
\hline \hline
\multicolumn{ 10}{c}{{\bf IC0167 - Generic Spectral Information on the Slit - 2006 Observation}} \\
\hline \hline
\multicolumn{ 1}{|c|}{{\bf ID}} & {\bf RA}    & {\bf Dec}   & {\bf $\Delta$y}                 & \multicolumn{ 1}{c}{\bf r}      & \multicolumn{ 1}{c}{\bf $\theta$} & \multicolumn{ 1}{c}{\bf v$_{obs}$} & \multicolumn{ 1}{c}{\bf $\sigma_{v_{obs}}$} & {\bf n$_{us}$} & {\bf n$_{lin}$} \\
\multicolumn{ 1}{|c|}{}         & {\bf [\dg]} & {\bf [\dg]} & \multicolumn{ 1}{c}{\bf [\dpr]} & \multicolumn{ 1}{c}{\bf [\dpr]} & \multicolumn{ 1}{c}{\bf [\dg]}    & \multicolumn{ 1}{c}{\bf [km/s]}    & \multicolumn{ 1}{c}{\bf [km/s]}             &                &                 \\
\hline
   {\bf 1} &   27.76827 &   21.92110 &          3 &         67 &        167 &       2832 &         36 &         17 &         18 \\
   {\bf 2} &   27.76503 &   21.91920 &          3 &         77 &        158 &       2749 &         54 &          8 &         14 \\
   {\bf 3} &   27.77588 &   21.92118 &          3 &         44 &       -177 &       2839 &         28 &         15 &         16 \\
   {\bf 4} &   27.77720 &   21.92236 &          3 &         45 &       -169 &       2954 &         54 &         12 &         16 \\
   {\bf 5} &   27.75487 &   21.89853 &          3 &        143 &        114 &     133706 &         14 &          5 &          5 \\
\hline \hline
\end{tabular}
\end{table*}

\begin{table}
 \caption{Correspondences between the slits observed in 2006 (ID06) and the
 slits observed in 2005 (ID05) for the galaxy IC0167.}
 \label{tblcorresp-1}
  \centering
   \begin{tabular}{ccccrr}
  \hline \hline

\multicolumn{ 6}{c}{{\bf IC0167 - Correspondences 2006 / 2005 }} \\
{\bf ID06} & {\bf ID05} & {\bf ID06} & {\bf ID05} & {\bf ID06} & {\bf ID05} \\
\hline
   {\bf 1} &         24 &   {\bf 15} &          6 &   {\bf 24} &          9 \\
   {\bf 2} &         25 &   {\bf 16} &         10 &   {\bf 25} &          8 \\
   {\bf 3} &         21 &   {\bf 17} &         13 &   {\bf 26} &         44 \\
   {\bf 4} &         20 &   {\bf 18} &         15 &   {\bf 27} &         43 \\
   {\bf 5} &         42 &   {\bf 19} &         16 &   {\bf 28} &         17 \\
   {\bf 7} &         26 &   {\bf 20} &         14 &   {\bf 30} &          4 \\
   {\bf 8} &         23 &   {\bf 21} &         12 &   {\bf 31} &          7 \\
   {\bf 9} &         22 &   {\bf 22} &         11 &   {\bf 34} &          2 \\
  {\bf 11} &         18 &   {\bf 23} &          5 &     {\bf } &            \\
\hline \hline
\end{tabular}
\end{table}

\clearpage
\begin{table*}
 \caption{Information about the spectra observed for NGC1042 in 2005.}
 \label{tblslit-3}
  \centering
\begin{tabular}{ccccrrrrcc}
\hline \hline
\multicolumn{ 10}{c}{{\bf NGC1042 - Generic Spectral Information on the Slit - 2005 Observation}} \\
\hline \hline
\multicolumn{ 1}{|c|}{{\bf ID}} & {\bf RA}    & {\bf Dec}   & {\bf $\Delta$y}                 & \multicolumn{ 1}{c}{\bf r}      & \multicolumn{ 1}{c}{\bf $\theta$} & \multicolumn{ 1}{c}{\bf v$_{obs}$} & \multicolumn{ 1}{c}{\bf $\sigma_{v_{obs}}$} & {\bf n$_{us}$} & {\bf n$_{lin}$} \\
\multicolumn{ 1}{|c|}{}         & {\bf [\dg]} & {\bf [\dg]} & \multicolumn{ 1}{c}{\bf [\dpr]} & \multicolumn{ 1}{c}{\bf [\dpr]} & \multicolumn{ 1}{c}{\bf [\dg]}    & \multicolumn{ 1}{c}{\bf [km/s]}    & \multicolumn{ 1}{c}{\bf [km/s]}             &                &                 \\
\hline
   {\bf 1} &   40.08535 &   -8.47102 &          5 &        165 &        -84 &       1334 &         17 &         13 &         14 \\
   {\bf 2} &   40.09769 &   -8.47767 &          5 &        181 &       -101 &       1400 &        100 &          8 &         12 \\
   {\bf 3} &   40.09000 &   -8.46668 &          5 &        143 &        -89 &       1350 &         46 &         14 &         16 \\
   {\bf 4} &   40.07942 &   -8.46420 &          5 &        150 &        -73 &          - &          - &          - &          - \\
   {\bf 5} &   40.08778 &   -8.46018 &          5 &        120 &        -81 &          - &          - &          - &          - \\
\hline \hline
\end{tabular}
\end{table*}

\begin{table*}
 \caption{Information about the spectra observed for NGC1042 in 2006.}
 \label{tblslit-4}
  \centering
\begin{tabular}{ccccrrrrcc}
\hline \hline
\multicolumn{ 10}{c}{{\bf NGC1042 - Generic Spectral Information on the Slit - 2006 Observation}} \\
\hline \hline
\multicolumn{ 1}{|c|}{{\bf ID}} & {\bf RA}    & {\bf Dec}   & {\bf $\Delta$y}                 & \multicolumn{ 1}{c}{\bf r}      & \multicolumn{ 1}{c}{\bf $\theta$} & \multicolumn{ 1}{c}{\bf v$_{obs}$} & \multicolumn{ 1}{c}{\bf $\sigma_{v_{obs}}$} & {\bf n$_{us}$} & {\bf n$_{lin}$} \\
\multicolumn{ 1}{|c|}{}         & {\bf [\dg]} & {\bf [\dg]} & \multicolumn{ 1}{c}{\bf [\dpr]} & \multicolumn{ 1}{c}{\bf [\dpr]} & \multicolumn{ 1}{c}{\bf [\dg]}    & \multicolumn{ 1}{c}{\bf [km/s]}    & \multicolumn{ 1}{c}{\bf [km/s]}             &                &                 \\
\hline
   {\bf 1} &   40.07691 &   -8.39319 &          4 &        179 &         50 &      18917 &         93 &          7 &         13 \\
   {\bf 2} &   40.11228 &   -8.41074 &          5 &        106 &        103 &       1320 &         23 &          9 &         12 \\
   {\bf 3} &   40.11674 &   -8.40818 &          3 &        123 &        107 &      31230 &        120 &          5 &          7 \\
   {\bf 4} &   40.08996 &   -8.40102 &          4 &        135 &         62 &       1394 &         23 &         18 &         20 \\
   {\bf 5} &   40.10812 &   -8.43035 &          4 &         33 &        144 &       1321 &         50 &         12 &         15 \\
\hline \hline
\end{tabular}
\end{table*}

\begin{table*}
 \contcaption{}
  \centering
\begin{tabular}{ccccrrrrcc}
\hline \hline
\multicolumn{ 10}{c}{{\bf NGC1042 - Generic Spectral Information on the Slit - 2006 Observation}} \\
\hline \hline
\multicolumn{ 1}{|c|}{{\bf ID}} & {\bf RA}    & {\bf Dec}   & {\bf $\Delta$y}                 & \multicolumn{ 1}{c}{\bf r}      & \multicolumn{ 1}{c}{\bf $\theta$} & \multicolumn{ 1}{c}{\bf v$_{obs}$} & \multicolumn{ 1}{c}{\bf $\sigma_{v_{obs}}$} & {\bf n$_{us}$} & {\bf n$_{lin}$} \\
\multicolumn{ 1}{|c|}{}         & {\bf [\dg]} & {\bf [\dg]} & \multicolumn{ 1}{c}{\bf [\dpr]} & \multicolumn{ 1}{c}{\bf [\dpr]} & \multicolumn{ 1}{c}{\bf [\dg]}    & \multicolumn{ 1}{c}{\bf [km/s]}    & \multicolumn{ 1}{c}{\bf [km/s]}             &                &                 \\
\hline
  {\bf 64} &   40.07985 &   -8.44607 &          3 &         91 &        -49 &       1395 &         56 &         11 &         14 \\
  {\bf 65} &   40.09874 &   -8.45494 &          3 &         88 &       -101 &       1373 &         25 &         16 &         19 \\
  {\bf 66} &   40.06881 &   -8.44617 &          3 &        126 &        -38 &       1421 &         32 &         12 &         16 \\
  {\bf 67} &   40.08476 &   -8.45375 &          3 &        102 &        -70 &       6602 &         55 &         11 &         18 \\
  {\bf 68} &   40.08262 &   -8.45569 &          3 &        113 &        -69 &       1356 &         39 &         12 &         17 \\
\hline \hline
\end{tabular}
\end{table*}

\begin{table}
 \caption{Correspondences between the slits observed in 2006 (ID06) and the
 slits observed in 2005 (ID05) for the galaxy NGC1042.}
 \label{tblcorresp-2}
  \centering
   \begin{tabular}{ccccrr}
  \hline \hline

\multicolumn{ 6}{c}{{\bf NGC1042 - Correspondences 2006 / 2005 }} \\
{\bf ID06} & {\bf ID05} & {\bf ID06} & {\bf ID05} & {\bf ID06} & {\bf ID05} \\
\hline
   {\bf 2} &         29 &   {\bf 15} &         19 &   {\bf 27} &          3 \\
   {\bf 3} &         31 &   {\bf 16} &         15 &   {\bf 29} & 5 (middle) \\
   {\bf 4} &         30 &   {\bf 17} &         20 &   {\bf 30} &          9 \\
   {\bf 5} &         21 &   {\bf 18} &         18 &   {\bf 32} &          4 \\
   {\bf 6} &         23 &   {\bf 19} &         16 &   {\bf 33} &          2 \\
   {\bf 8} &         28 &   {\bf 20} &         14 &   {\bf 34} &          1 \\
  {\bf 10} &         22 &   {\bf 21} &         12 &   {\bf 35} &         37 \\
  {\bf 11} &         35 &   {\bf 22} &         13 &   {\bf 47} &         26 \\
  {\bf 13} &         24 &   {\bf 24} &          7 &   {\bf 70} & 5 (bottom) \\
\hline \hline
\end{tabular}
\end{table}

\clearpage
\begin{table*}
 \caption{Information about the spectra observed for NGC6907 in 2005.}
 \label{tblslit-5}
  \centering
\begin{tabular}{ccccrrrrcc}
\hline \hline
\multicolumn{ 10}{c}{{\bf NGC6907 - Generic Spectral Information on the Slit - 2005 Observation}} \\
\hline \hline
\multicolumn{ 1}{|c|}{{\bf ID}} & {\bf RA}    & {\bf Dec}   & {\bf $\Delta$y}                 & \multicolumn{ 1}{c}{\bf r}      & \multicolumn{ 1}{c}{\bf $\theta$} & \multicolumn{ 1}{c}{\bf v$_{obs}$} & \multicolumn{ 1}{c}{\bf $\sigma_{v_{obs}}$} & {\bf n$_{us}$} & {\bf n$_{lin}$} \\
\multicolumn{ 1}{|c|}{}         & {\bf [\dg]} & {\bf [\dg]} & \multicolumn{ 1}{c}{\bf [\dpr]} & \multicolumn{ 1}{c}{\bf [\dpr]} & \multicolumn{ 1}{c}{\bf [\dg]}    & \multicolumn{ 1}{c}{\bf [km/s]}    & \multicolumn{ 1}{c}{\bf [km/s]}             &                &                 \\
\hline
   {\bf 1} &  306.30016 &  -24.76257 &          5 &        223 &        137 &       6280 &        180 &          4 &          9 \\
   {\bf 3} &  306.30690 &  -24.78402 &          5 &        137 &       -176 &      29817 &         97 &          2 &          3 \\
   {\bf 4} &  306.30758 &  -24.78843 &          5 &        125 &       -163 &       3490 &        140 &          5 &         10 \\
   {\bf 5} &  306.30240 &  -24.79045 &          5 &        108 &       -168 &      16327 &         96 &          5 &          7 \\
   {\bf 6} &  306.29779 &  -24.78856 &          5 &        106 &        173 &      14758 &         98 &          7 &         12 \\
\hline \hline
\end{tabular}
\end{table*}

\begin{table*}
 \caption{Information about the spectra observed for NGC6907 in 2006.}
 \label{tblslit-6}
  \centering
\begin{tabular}{ccccrrrrcc}
\hline \hline
\multicolumn{ 10}{c}{{\bf NGC6907 - Generic Spectral Information on the Slit - 2006 Observation}} \\
\hline \hline
\multicolumn{ 1}{|c|}{{\bf ID}} & {\bf RA}    & {\bf Dec}   & {\bf $\Delta$y}                 & \multicolumn{ 1}{c}{\bf r}      & \multicolumn{ 1}{c}{\bf $\theta$} & \multicolumn{ 1}{c}{\bf v$_{obs}$} & \multicolumn{ 1}{c}{\bf $\sigma_{v_{obs}}$} & {\bf n$_{us}$} & {\bf n$_{lin}$} \\
\multicolumn{ 1}{|c|}{}         & {\bf [\dg]} & {\bf [\dg]} & \multicolumn{ 1}{c}{\bf [\dpr]} & \multicolumn{ 1}{c}{\bf [\dpr]} & \multicolumn{ 1}{c}{\bf [\dg]}    & \multicolumn{ 1}{c}{\bf [km/s]}    & \multicolumn{ 1}{c}{\bf [km/s]}             &                &                 \\
\hline
   {\bf 1} &  306.24745 &  -24.81042 &          3 &        112 &         61 &       3261 &         46 &         12 &         16 \\
   {\bf 2} &  306.26098 &  -24.81597 &          5 &         60 &         40 &       3374 &         42 &         16 &         17 \\
   {\bf 3} &  306.24993 &  -24.80975 &          4 &        104 &         62 &       3318 &         41 &         15 &         17 \\
   {\bf 4} &  306.24475 &  -24.83186 &          3 &        136 &         17 &      64208 &         53 &          1 &          3 \\
   {\bf 5} &  306.26041 &  -24.81775 &          5 &         64 &         33 &       3342 &         40 &         14 &         18 \\
\hline \hline
\end{tabular}
\end{table*}

\begin{table}
 \caption{Correspondences between the slits observed in 2006 (ID06) and the
 slits observed in 2005 (ID05) for the galaxy NGC6907.}
 \label{tblcorresp-3}
  \centering
   \begin{tabular}{ccccrr}
  \hline \hline

\multicolumn{ 6}{c}{{\bf NGC6907 - Correspondences 2006 / 2005 }} \\
{\bf ID06} & {\bf ID05} & {\bf ID06} & {\bf ID05} & {\bf ID06} & {\bf ID05} \\
\hline
   {\bf 1} &         34 &   {\bf 15} &         27 &   {\bf 26} &          8 \\
   {\bf 2} &         28 &   {\bf 16} &         16 &   {\bf 27} &         10 \\
   {\bf 5} &         29 &   {\bf 17} &         22 &   {\bf 29} &         12 \\
   {\bf 7} &         26 &   {\bf 20} &         14 &   {\bf 35} &          3 \\
   {\bf 8} &         25 &   {\bf 21} &         18 &   {\bf 53} &         16 \\
   {\bf 9} &         23 &   {\bf 23} &         15 &   {\bf 59} &          7 \\
  {\bf 10} &         24 &   {\bf 24} &         19 &            &            \\
  {\bf 12} &         35 &   {\bf 25} &         11 &            &            \\
\hline \hline
\end{tabular}
\end{table}

\clearpage
\section{Intrinsic Extinctions}

\par The following tables in this appendix contain the intrinsic extinctions
calculated using the Balmer lines observed in the spectrum
associated to each slit. They were evaluated using the
H$\gamma$/H$\beta$ ratio for the 2005 observations and the
H$\alpha$/H$\beta$ ratio for the 2006 observations. ID is the
identification number of the slit and the E$_{B-V}$ is the intrinsic
extinction.

\begin{table}
 \caption{Intrinsic extinctions for the regions observed in 2005 for IC0167.}
 \label{tblext-1}
  \centering
\begin{tabular}{cccccccc}
\hline \hline
                                          \multicolumn{ 8}{c}{{\it {\bf IC0167 - Extinction - 2005}}} \\
\hline \hline
{\it {\bf ID}} & {\bf E$_{B-V}$} & {\bf ID} & {\bf E$_{B-V}$} & {\bf ID} & {\bf E$_{B-V}$} & {\bf ID} & {\bf E$_{B-V}$} \\
\hline
   {\bf 2}  &       0.16 &   {\bf 11} &       0.02 &   {\bf 17} &       0.09 &   {\bf 26} &       0.09 \\
   {\bf 5}  &       0.02 &   {\bf 12} &       0.05 &   {\bf 18} &       0.07 &   {\bf 28} &       0.09 \\
   {\bf 8}  &       0.15 &   {\bf 13} &       0.10 &   {\bf 19} &       0.09 &   {\bf 43} &       0.11 \\
   {\bf 9}  &       0.04 &   {\bf 14} &       0.08 &   {\bf 21} &       0.12 &            &            \\
   {\bf 10} &       0.10 &   {\bf 15} &       0.06 &   {\bf 24} &       0.07 &            &            \\
\hline \hline
\end{tabular}
\end{table}

\begin{table}
 \caption{Intrinsic extinctions for the regions observed in 2006 for IC0167.}
 \label{tblext-2}
  \centering
\begin{tabular}{cccccccc}
\hline \hline
                                          \multicolumn{ 8}{c}{{\it {\bf IC0167 - Extinction - 2006}}} \\
\hline \hline
{\it {\bf ID}} & {\bf E$_{B-V}$} & {\bf ID} & {\bf E$_{B-V}$} & {\bf ID} & {\bf E$_{B-V}$} & {\bf ID} & {\bf E$_{B-V}$} \\
\hline
   {\bf 1}  &       0.02 &   {\bf 15} &       0.04 &   {\bf 24} &       0.01 &   {\bf 46} &       0.13 \\
   {\bf 3}  &       0.16 &   {\bf 16} &       0.01 &   {\bf 25} &       0.20 &   {\bf 50} &       0.04 \\
   {\bf 7}  &       0.16 &   {\bf 18} &       0.07 &   {\bf 27} &       0.09 &   {\bf 51} &       0.14 \\
   {\bf 8}  &       0.09 &   {\bf 20} &       0.02 &   {\bf 28} &       0.11 &   {\bf 55} &       0.16 \\
   {\bf 9}  &       0.19 &   {\bf 21} &       0.01 &   {\bf 43} &       0.05 &            &            \\
   {\bf 11} &       0.18 &   {\bf 22} &       0.12 &   {\bf 44} &       0.09 &            &            \\
\hline \hline
\end{tabular}
\end{table}

\begin{table}
 \caption{Intrinsic extinctions for the regions observed in 2005 for NGC1042.}
 \label{tblext-3}
  \centering
\begin{tabular}{cccccccc}
\hline \hline
                                          \multicolumn{ 8}{c}{{\it {\bf NGC1042 - Extinction - 2005}}} \\
\hline \hline
{\it {\bf ID}} & {\bf E$_{B-V}$} & {\bf ID} & {\bf E$_{B-V}$} & {\bf ID} & {\bf E$_{B-V}$} & {\bf ID} & {\bf E$_{B-V}$} \\
\hline
   {\bf 1} &       0.21 &   {\bf 10} &       0.18 &   {\bf 20} &       0.16 &   {\bf 28} &       0.16 \\
   {\bf 3} &       0.11 &   {\bf 12} &       0.09 &   {\bf 21} &       0.10 &   {\bf 30} &       0.14 \\
   {\bf 7} &       0.16 &   {\bf 16} &       0.16 &   {\bf 22} &       0.18 &   {\bf 35} &       0.19 \\
   {\bf 9} &       0.14 &   {\bf 18} &       0.12 &   {\bf 25} &       0.15 &   {\bf 37} &       0.18 \\
\hline \hline
\end{tabular}
\end{table}

\begin{table}
 \caption{Intrinsic extinctions for the regions observed in 2006 for NGC1042.}
 \label{tblext-4}
  \centering
\begin{tabular}{cccccccc}
\hline \hline
                                          \multicolumn{ 8}{c}{{\it {\bf NGC1042 - Extinction - 2006}}} \\
\hline \hline
{\it {\bf ID}} & {\bf E$_{B-V}$} & {\bf ID} & {\bf E$_{B-V}$} & {\bf ID} & {\bf E$_{B-V}$} & {\bf ID} & {\bf E$_{B-V}$} \\
\hline
   {\bf 2} &       0.10 &   {\bf 17} &       0.13 &   {\bf 34} &       0.08 &   {\bf 56} &       0.16 \\
   {\bf 4} &       0.10 &   {\bf 18} &       0.23 &   {\bf 43} &       0.13 &   {\bf 57} &       0.13 \\
   {\bf 5} &       0.21 &   {\bf 19} &       0.08 &   {\bf 45} &       0.09 &   {\bf 59} &       0.06 \\
   {\bf 6} &       0.13 &   {\bf 21} &       0.16 &   {\bf 48} &       0.27 &   {\bf 61} &       0.17 \\
   {\bf 8} &       0.15 &   {\bf 22} &       0.12 &   {\bf 49} &       0.17 &   {\bf 62} &       0.16 \\
   {\bf 9} &       0.25 &   {\bf 23} &       0.15 &   {\bf 50} &       0.07 &   {\bf 64} &       0.22 \\
  {\bf 10} &       0.12 &   {\bf 24} &       0.12 &   {\bf 51} &       0.25 &   {\bf 65} &       0.23 \\
  {\bf 12} &       0.22 &   {\bf 25} &       0.29 &   {\bf 53} &       0.17 &   {\bf 68} &       0.13 \\
  {\bf 14} &       0.20 &   {\bf 27} &       0.16 &   {\bf 54} &       0.23 &   {\bf 70} &       0.20 \\
  {\bf 15} &       0.22 &   {\bf 32} &       0.14 &   {\bf 55} &       0.11 &            &            \\
\hline \hline
\end{tabular}
\end{table}

\begin{table}
 \caption{Intrinsic extinctions for the regions observed in 2005 for NGC6907.}
 \label{tblext-5}
  \centering
\begin{tabular}{cccccccc}
\hline \hline
                                          \multicolumn{ 8}{c}{{\it {\bf NGC6907 - Extinction - 2005}}} \\
\hline \hline
{\it {\bf ID}} & {\bf E$_{B-V}$} & {\bf ID} & {\bf E$_{B-V}$} & {\bf ID} & {\bf E$_{B-V}$} & {\bf ID} & {\bf E$_{B-V}$} \\
\hline
   {\bf 8} &       0.50 &   {\bf 17} &       0.47 &   {\bf 24} &       0.51 &   {\bf 33} &       0.48 \\
  {\bf 10} &       0.38 &   {\bf 18} &       0.60 &   {\bf 25} &       0.49 &   {\bf 34} &       0.36 \\
  {\bf 11} &       0.42 &   {\bf 19} &       0.56 &   {\bf 26} &       0.39 &   {\bf 35} &       0.48 \\
  {\bf 12} &       0.50 &   {\bf 20} &       0.60 &   {\bf 27} &       0.41 &            &            \\
  {\bf 14} &       0.42 &   {\bf 21} &       0.48 &   {\bf 28} &       0.48 &            &            \\
  {\bf 15} &       0.40 &   {\bf 22} &       0.46 &   {\bf 29} &       0.25 &            &            \\
  {\bf 16} &       0.32 &   {\bf 23} &       0.77 &   {\bf 31} &       0.26 &            &            \\
\hline \hline
\end{tabular}
\end{table}

\begin{table}
 \caption{Intrinsic extinctions for the regions observed in 2006 for NGC6907.}
 \label{tblext-6}
  \centering
\begin{tabular}{cccccccc}
\hline \hline
                                          \multicolumn{ 8}{c}{{\it {\bf NGC6907 - Extinction - 2006}}} \\
\hline \hline
{\it {\bf ID}} & {\bf E$_{B-V}$} & {\bf ID} & {\bf E$_{B-V}$} & {\bf ID} & {\bf E$_{B-V}$} & {\bf ID} & {\bf E$_{B-V}$} \\
\hline
   {\bf 1} &       0.33 &   {\bf 15} &       0.25 &   {\bf 27} &       0.54 &   {\bf 50} &       0.55 \\
   {\bf 2} &       0.34 &   {\bf 16} &       0.47 &   {\bf 28} &       0.27 &   {\bf 51} &       0.43 \\
   {\bf 3} &       0.48 &   {\bf 17} &       0.52 &   {\bf 29} &       0.52 &   {\bf 53} &       0.55 \\
   {\bf 5} &       0.15 &   {\bf 20} &       0.54 &   {\bf 37} &       0.27 &   {\bf 54} &       0.54 \\
   {\bf 7} &       0.40 &   {\bf 21} &       0.61 &   {\bf 40} &       0.38 &   {\bf 55} &       0.57 \\
   {\bf 8} &       0.55 &   {\bf 22} &       0.36 &   {\bf 45} &       0.25 &   {\bf 57} &       0.30 \\
   {\bf 9} &       0.28 &   {\bf 23} &       0.79 &   {\bf 46} &       0.37 &   {\bf 61} &       0.42 \\
  {\bf 10} &       0.15 &   {\bf 24} &       0.55 &   {\bf 47} &       0.12 &            &            \\
  {\bf 12} &       0.35 &   {\bf 25} &       0.10 &   {\bf 48} &       0.41 &            &            \\
  {\bf 14} &       0.36 &   {\bf 26} &       0.29 &   {\bf 49} &       0.59 &            &            \\
\hline \hline
\end{tabular}
\end{table}

\clearpage
\section{Spectral Lines Fluxes}

\par The tables in this appendix show the mean flux measured for each
spectral line used to evaluate the metallicities, according to the
different procedures mentioned on the paper. Slits flagged with a
"$^\dag$" were removed from our analysis due the underlying Balmer
absorption.

\begin{table*}
 \caption{Fluxes of the spectral lines observed in IC0167 using the
  slit identified by the year of the observation and its ID. Slits flagged with a "$^\dag$" were removed
  from our analysis due the underlying Balmer absorption.}
 \label{tblflx-1}
  \centering
\begin{tabular}{crrrrrrrr}
\hline \hline
\multicolumn{ 9}{c}{{\bf IC0167 - Fluxes [10$^{-17}$erg/cm$^{2}$/s/$\AA$]}} \\
\hline \hline
{\bf Slit} & \multicolumn{ 1}{c}{\bf [O~{\sc ii}]} & \multicolumn{ 1}{c}{\bf H$\beta$} & \multicolumn{ 1}{c}{\bf [O~{\sc iii}]} & \multicolumn{ 1}{c}{\bf [O~{\sc iii}]} & \multicolumn{ 1}{c}{\bf [N~{\sc ii}]} & \multicolumn{ 1}{c}{\bf H$\alpha$} & \multicolumn{ 1}{c}{\bf [N~{\sc ii}]} & \multicolumn{ 1}{c}{\bf [Ar~{\sc iii}]} \\
{\bf Year$\_$ID} & {\bf (3727$\AA$)} & {\bf (4861$\AA$)} & {\bf (4959$\AA$)} & {\bf (5007$\AA$)} & {\bf (6548$\AA$)} & {\bf (6563$\AA$)} & {\bf (6584$\AA$)} & {\bf (7136$\AA$)} \\
\hline
     05$\_$05$^\dag$ &                    - &    7.01$\pm$0.24 &                    - &                    - &                    - &                    - &                    - &                    - \\
     05$\_$08$^\dag$ &                    - &    217.0$\pm$2.7 &    223.4$\pm$2.2 &    629.1$\pm$7.5 &                    - &                    - &                    - &                    - \\
            05$\_$09 &       316$\pm$46 &  104.20$\pm$0.55 &   92.88$\pm$0.14 &    263.5$\pm$3.4 &                    - &                    - &                    - &                    - \\
     05$\_$10$^\dag$ &     70.7$\pm$6.6 &     26.9$\pm$1.2 &   31.74$\pm$0.65 &  109.50$\pm$0.05 &                    - &                    - &                    - &                    - \\
            05$\_$11 &     1620$\pm$230 &       865$\pm$10 &      1392$\pm$16 &      3763$\pm$34 &                    - &                    - &                    - &                    - \\
            05$\_$12 &       209$\pm$55 &   47.25$\pm$0.09 &   37.67$\pm$0.37 &  101.20$\pm$0.40 &                    - &                    - &                    - &                    - \\
     05$\_$13$^\dag$ &                    - &      6.1$\pm$1.2 &    1.56$\pm$0.36 &    5.98$\pm$0.57 &                    - &                    - &                    - &                    - \\
     05$\_$14$^\dag$ &                    - &   11.49$\pm$0.60 &   12.46$\pm$0.27 &   39.98$\pm$0.09 &    0.56$\pm$0.22 &                    - &                    - &                    - \\
            05$\_$15 &     3430$\pm$420 &      1380$\pm$14 &    621.3$\pm$7.1 &      1830$\pm$26 &                    - &                    - &                    - &                    - \\
     05$\_$17$^\dag$ &      250$\pm$130 &   68.50$\pm$0.37 &   25.63$\pm$0.81 &   80.26$\pm$0.87 &                    - &                    - &                    - &                    - \\
            05$\_$18 &       513$\pm$51 &    205.5$\pm$1.4 &  100.40$\pm$0.15 &    298.1$\pm$3.8 &                    - &                    - &                    - &                    - \\
            05$\_$19 &       154$\pm$22 &   29.47$\pm$0.05 &   24.00$\pm$0.30 &   75.07$\pm$0.02 &                    - &                    - &                    - &                    - \\
            05$\_$21 &    193.1$\pm$6.3 &   94.37$\pm$0.75 &   55.40$\pm$0.89 &  169.30$\pm$0.10 &                    - &                    - &                    - &                    - \\
            05$\_$24 &       218$\pm$20 &   81.19$\pm$0.19 &   81.93$\pm$0.62 &    243.2$\pm$2.7 &                    - &                    - &                    - &                    - \\
            05$\_$26 &       218$\pm$10 &     68.5$\pm$1.1 &   66.11$\pm$0.59 &    199.9$\pm$2.3 &                    - &                    - &                    - &                    - \\
            05$\_$28 &       424$\pm$17 &    173.2$\pm$2.2 &  180.10$\pm$0.15 &    526.2$\pm$1.1 &                    - &                    - &                    - &                    - \\
     05$\_$43$^\dag$ &        29$\pm$10 &     16.6$\pm$1.3 &   15.19$\pm$0.70 &     54.1$\pm$2.0 &                    - &                    - &                    - &                    - \\
            06$\_$01 &                    - &   44.86$\pm$0.01 &   49.61$\pm$0.21 &  149.20$\pm$0.90 &    0.79$\pm$0.13 &  116.10$\pm$0.85 &    4.22$\pm$0.10 &                    - \\
            06$\_$03 &                    - &     47.9$\pm$1.2 &   22.06$\pm$0.17 &     61.0$\pm$3.2 &    3.70$\pm$0.20 &  135.30$\pm$0.50 &   16.70$\pm$0.01 &                    - \\
            06$\_$07 &                    - &     73.8$\pm$1.0 &   65.03$\pm$0.60 &    195.8$\pm$1.8 &    4.12$\pm$0.36 &  167.90$\pm$0.85 &     13.8$\pm$2.3 &    5.03$\pm$0.67 \\
            06$\_$08 &                    - &   16.09$\pm$0.80 &    8.47$\pm$0.51 &   30.81$\pm$0.79 &    0.86$\pm$0.27 &   40.80$\pm$0.03 &      4.5$\pm$2.3 &                    - \\
            06$\_$09 &                    - &  251.60$\pm$0.90 &    192.1$\pm$1.0 &    583.3$\pm$5.0 &   13.67$\pm$0.04 &    551.7$\pm$3.9 &   44.46$\pm$0.32 &   10.26$\pm$0.08 \\
            06$\_$11 &                    - &  212.40$\pm$0.60 &  112.20$\pm$0.15 &    308.8$\pm$1.3 &   26.65$\pm$0.07 &    592.1$\pm$3.6 &   82.91$\pm$0.32 &   11.70$\pm$0.25 \\
            06$\_$15 &                    - &     32.9$\pm$1.1 &   41.19$\pm$0.11 &  120.00$\pm$0.10 &    0.63$\pm$0.06 &   85.24$\pm$0.74 &    4.01$\pm$0.03 &    1.57$\pm$0.28 \\
            06$\_$16 &                    - &   19.61$\pm$0.23 &   20.14$\pm$0.48 &   52.22$\pm$0.46 &    0.43$\pm$0.23 &   54.34$\pm$0.01 &    3.41$\pm$0.39 &    1.35$\pm$0.04 \\
            06$\_$18 &                    - &   1076.0$\pm$9.0 &    498.5$\pm$2.2 &   1559.0$\pm$7.5 &    135.7$\pm$1.5 &      2981$\pm$19 &    421.5$\pm$4.1 &   66.79$\pm$0.16 \\
            06$\_$20 &                    - &   11.36$\pm$0.42 &    6.29$\pm$0.97 &   22.29$\pm$0.77 &    1.07$\pm$0.34 &   31.40$\pm$0.35 &      2.5$\pm$1.6 &                    - \\
            06$\_$21 &                    - &   32.45$\pm$0.06 &   22.31$\pm$0.05 &   71.05$\pm$0.11 &    0.82$\pm$0.16 &   89.62$\pm$0.39 &    5.67$\pm$0.07 &    1.97$\pm$0.15 \\
            06$\_$22 &                    - &  817.00$\pm$0.65 &   1326.0$\pm$2.5 &   4099.0$\pm$7.0 &   26.10$\pm$0.01 &   1883.0$\pm$8.0 &   82.24$\pm$0.46 &   47.00$\pm$0.13 \\
            06$\_$24 &                    - &   69.48$\pm$0.44 &   55.14$\pm$0.25 &    182.2$\pm$1.7 &    3.28$\pm$0.16 &    191.8$\pm$2.0 &   11.90$\pm$0.01 &    3.94$\pm$0.01 \\
            06$\_$25 &                    - &   175.5$\pm$1.35 &   167.1$\pm$1.25 &      492.7$\pm$1 &    5.71$\pm$0.15 &    359.6$\pm$0.8 &   19.73$\pm$0.09 &    7.08$\pm$0.22 \\
            06$\_$27 &                    - &   16.45$\pm$0.47 &   12.26$\pm$0.25 &   37.29$\pm$0.99 &    0.91$\pm$0.07 &   32.14$\pm$0.05 &    1.21$\pm$0.50 &    1.29$\pm$0.28 \\
            06$\_$28 &                    - &   63.55$\pm$0.08 &     29.1$\pm$6.5 &     78.2$\pm$4.2 &    8.53$\pm$0.91 &    177.0$\pm$1.8 &   24.81$\pm$0.06 &    1.16$\pm$0.08 \\
            06$\_$43 &                    - &   31.26$\pm$0.19 &   33.29$\pm$0.71 &   90.94$\pm$0.23 &    1.55$\pm$0.12 &   86.49$\pm$0.65 &    6.96$\pm$0.10 &    1.78$\pm$0.07 \\
            06$\_$44 &                    - &   77.03$\pm$2.53 &   68.82$\pm$0.39 &   197.7$\pm$1.35 &    8.38$\pm$2.07 &   218.6$\pm$2.45 &    16.5$\pm$0.81 &    0.91$\pm$0.35 \\
            06$\_$46 &                    - &    4.67$\pm$0.38 &    2.15$\pm$0.12 &   10.72$\pm$0.47 &    0.37$\pm$0.18 &   13.73$\pm$0.19 &    1.16$\pm$0.13 &    0.33$\pm$0.16 \\
            06$\_$50 &                    - &   12.36$\pm$0.11 &    7.89$\pm$0.07 &   28.01$\pm$0.35 &    1.46$\pm$0.14 &   34.44$\pm$0.17 &    2.99$\pm$0.20 &    0.13$\pm$0.07 \\
            06$\_$51 &                    - &   34.82$\pm$0.37 &    6.32$\pm$0.65 &   24.97$\pm$0.62 &    3.72$\pm$0.22 &   96.77$\pm$0.58 &   11.56$\pm$0.33 &    1.18$\pm$0.11 \\
            06$\_$55 &                    - &   27.78$\pm$0.14 &   21.81$\pm$0.33 &   58.21$\pm$0.16 &    1.67$\pm$0.27 &   77.33$\pm$0.59 &    7.44$\pm$0.51 &                    - \\
\hline \hline
\end{tabular}
\end{table*}

\begin{table*}
 \caption{Fluxes of the spectral lines observed in NGC1042 using the
  slit identified by the year of the observation and its ID. Slits flagged with a "$^\dag$" were removed
  from our analysis due the underlying Balmer absorption.}
 \label{tblflx-2}
  \centering
\begin{tabular}{crrrrrrrr}
\hline \hline
\multicolumn{ 9}{c}{{\bf NGC1042 - Fluxes [10$^{-17}$erg/cm$^{2}$/s/$\AA$]}} \\
\hline \hline
{\bf Slit} & \multicolumn{ 1}{c}{\bf [O~{\sc ii}]} & \multicolumn{ 1}{c}{\bf H$\beta$} & \multicolumn{ 1}{c}{\bf [O~{\sc iii}]} & \multicolumn{ 1}{c}{\bf [O~{\sc iii}]} & \multicolumn{ 1}{c}{\bf [N~{\sc ii}]} & \multicolumn{ 1}{c}{\bf H$\alpha$} & \multicolumn{ 1}{c}{\bf [N~{\sc ii}]} & \multicolumn{ 1}{c}{\bf [Ar~{\sc iii}]} \\
{\bf Year$\_$ID} & {\bf (3727$\AA$)} & {\bf (4861$\AA$)} & {\bf (4959$\AA$)} & {\bf (5007$\AA$)} & {\bf (6548$\AA$)} & {\bf (6563$\AA$)} & {\bf (6584$\AA$)} & {\bf (7136$\AA$)} \\
\hline
         05\_01 &      339$\pm$70 &  92.14$\pm$0.67 &   124.2$\pm$1.6 &   349.9$\pm$8.3 &               - &               - &               - &               - \\
         05\_03 &     1218$\pm$99 &   475.1$\pm$7.9 &   400.3$\pm$6.2 &     1193$\pm$18 &               - &               - &               - &               - \\
         05\_07 &      524$\pm$62 &   206.4$\pm$5.2 &   127.6$\pm$1.3 &   378.9$\pm$2.0 &               - &               - &               - &               - \\
         05\_09 &               - &  90.33$\pm$0.86 &     7.4$\pm$2.3 &    10.4$\pm$1.5 &    19.7$\pm$1.3 &   248.1$\pm$2.5 &  67.07$\pm$0.04 &               - \\
         05\_10 &               - &     1510$\pm$20 &   385.0$\pm$3.0 &  1136.0$\pm$9.2 &   339.8$\pm$3.3 &     4162$\pm$18 &     1039$\pm$11 &               - \\
         05\_12 &      452$\pm$43 & 231.60$\pm$0.78 &    36.8$\pm$2.1 &  90.43$\pm$0.03 &               - &               - &               - &               - \\
         05\_16 &      191$\pm$12 &   266.6$\pm$1.7 &     4.4$\pm$1.8 &    15.0$\pm$2.8 &               - &               - &               - &               - \\
         05\_18 &               - & 140.20$\pm$0.99 &     3.9$\pm$1.1 &    11.8$\pm$7.6 &  39.68$\pm$0.56 &   418.4$\pm$4.7 &               - &               - \\
         05\_20 &    1231$\pm$140 &   586.4$\pm$3.6 & 129.20$\pm$0.57 &   356.5$\pm$4.4 &               - &               - &               - &               - \\
         05\_21 &    56.4$\pm$1.0 &   131.0$\pm$6.0 &     3.8$\pm$1.9 &    11.3$\pm$2.3 &               - &               - &               - &               - \\
         05\_22 &     1351$\pm$97 &      977$\pm$16 &   136.7$\pm$1.7 &   378.4$\pm$6.2 &               - &               - &               - &               - \\
         05\_25 &      148$\pm$28 &    65.4$\pm$2.2 &     9.8$\pm$4.3 &    23.1$\pm$2.7 &               - &               - &               - &               - \\
  05\_28$^\dag$ &               - &    18.7$\pm$5.0 &     4.2$\pm$1.1 &     5.7$\pm$1.4 &   3.34$\pm$0.38 &  41.91$\pm$0.75 &  10.29$\pm$0.44 &               - \\
  05\_30$^\dag$ &               - &   287.4$\pm$4.9 & 102.50$\pm$0.85 &   254.4$\pm$2.8 &  50.02$\pm$0.65 &   745.1$\pm$8.0 &   150.9$\pm$1.8 &               - \\
         05\_35 &       35$\pm$10 &  14.80$\pm$0.91 &     2.6$\pm$1.1 &    12.6$\pm$1.0 &               - &               - &               - &               - \\
  05\_37$^\dag$ &      154$\pm$32 &  94.06$\pm$0.39 &     8.8$\pm$6.4 &    38.5$\pm$3.0 &               - &               - &               - &               - \\
         06\_02 &               - &     8.5$\pm$1.9 &               - &     2.8$\pm$1.8 &   1.66$\pm$0.83 &  33.02$\pm$0.64 &   7.33$\pm$0.71 &               - \\
         06\_04 &               - &   145.2$\pm$2.0 &  58.14$\pm$0.01 &   163.1$\pm$3.0 &  23.78$\pm$0.19 &   402.8$\pm$7.7 &    72.9$\pm$1.2 &   9.84$\pm$0.16 \\
         06\_05 &               - &   129.0$\pm$2.9 &   8.02$\pm$0.24 &    12.7$\pm$3.2 &  36.62$\pm$0.57 &   363.1$\pm$2.2 & 116.60$\pm$0.42 &   2.52$\pm$0.85 \\
         06\_06 &               - &    21.3$\pm$5.5 &    2.35$\pm$1.0 &   1.26$\pm$0.48 &     3.8$\pm$1.0 &  52.41$\pm$0.19 &  15.51$\pm$0.15 &               - \\
         06\_08 &               - &  10.16$\pm$0.57 &     2.9$\pm$1.2 &   4.91$\pm$0.96 &   2.77$\pm$0.62 &  37.50$\pm$0.86 &   8.05$\pm$0.74 &   0.60$\pm$0.07 \\
         06\_09 &               - & 121.10$\pm$0.57 &    21.8$\pm$1.8 &  70.89$\pm$0.23 &  31.10$\pm$0.01 &   338.4$\pm$4.5 &  90.42$\pm$0.91 &               - \\
         06\_10 &               - &   334.5$\pm$6.1 &  48.24$\pm$0.52 &   136.9$\pm$1.1 &  75.96$\pm$0.92 &      927$\pm$16 &   231.3$\pm$4.2 &               - \\
         06\_12 &               - &   7.77$\pm$0.58 &               - &               - &               - &  15.70$\pm$0.06 &   3.00$\pm$0.29 &               - \\
  06\_14$^\dag$ &               - &    27.6$\pm$2.8 &               - &    13.5$\pm$2.9 &   6.58$\pm$0.61 &    65.4$\pm$1.9 &  17.66$\pm$0.49 &               - \\
         06\_15 &               - & 133.80$\pm$0.42 &   1.25$\pm$0.87 &    16.3$\pm$1.4 &  37.53$\pm$0.09 & 373.40$\pm$0.28 &   112.0$\pm$1.1 &               - \\
         06\_17 &               - &   247.8$\pm$3.0 &  54.43$\pm$0.02 &    173$\pm$2.55 &  51.05$\pm$0.42 &   688.0$\pm$8.2 &   167.9$\pm$1.9 &               - \\
         06\_18 &               - &  89.25$\pm$0.62 &   8.47$\pm$0.25 &  11.76$\pm$0.91 &  23.84$\pm$0.23 &   248.9$\pm$2.0 &  73.23$\pm$0.44 &               - \\
         06\_19 &               - &    93.9$\pm$1.1 &   2.82$\pm$0.89 &   4.57$\pm$0.87 &  22.85$\pm$0.61 &   260.2$\pm$1.1 &  67.08$\pm$0.06 &     1.5$\pm$0.4 \\
         06\_21 &               - & 115.90$\pm$0.14 &  16.81$\pm$0.60 &  55.62$\pm$0.45 &  34.72$\pm$0.24 &   322.2$\pm$3.7 & 103.60$\pm$0.28 &   4.33$\pm$0.25 \\
         06\_22 &               - &  10.28$\pm$0.66 &   1.99$\pm$0.70 &   2.91$\pm$0.33 &   2.13$\pm$0.93 &  25.74$\pm$0.38 &   9.53$\pm$0.07 &               - \\
         06\_23 &               - & 104.40$\pm$0.28 &    27.1$\pm$2.6 &    84.9$\pm$3.6 &  13.41$\pm$0.33 & 235.20$\pm$0.49 &  46.66$\pm$0.16 &               - \\
         06\_24 &               - & 123.10$\pm$0.07 &  70.26$\pm$0.92 &   220.1$\pm$1.6 &  20.63$\pm$0.01 &   341.8$\pm$6.0 &    63.0$\pm$1.0 &   6.57$\pm$0.09 \\
         06\_25 &               - &  10.45$\pm$0.08 &     3.0$\pm$1.9 &  14.05$\pm$0.62 &   2.66$\pm$0.56 &  29.18$\pm$0.16 &   5.33$\pm$0.05 &               - \\
         06\_27 &               - &   252.9$\pm$3.5 &   206.6$\pm$3.0 &   609.4$\pm$8.4 &  17.60$\pm$0.04 &   480.4$\pm$4.0 &  55.77$\pm$0.18 &  11.32$\pm$0.36 \\
  06\_32$^\dag$ &               - &     4.2$\pm$2.7 &               - &  15.43$\pm$0.04 &               - &  12.57$\pm$0.33 &   1.47$\pm$0.53 &               - \\
         06\_34 &               - &  25.66$\pm$0.14 &  34.56$\pm$0.02 &  106.7$\pm$0.01 &   2.23$\pm$0.42 &    71.1$\pm$1.1 &   5.98$\pm$0.07 &   1.22$\pm$0.14 \\
         06\_43 &               - &   1.24$\pm$0.63 &               - &               - &   0.21$\pm$0.33 &   0.34$\pm$0.08 &               - &               - \\
         06\_45 &               - &   126.1$\pm$1.4 &  28.38$\pm$0.33 &  86.23$\pm$0.64 &  33.06$\pm$0.11 &   349.5$\pm$4.8 &    94.0$\pm$1.3 &               - \\
         06\_48 &               - &  53.79$\pm$0.86 &     7.3$\pm$3.5 &    11.5$\pm$7.0 &  13.49$\pm$0.18 & 150.40$\pm$0.35 &  44.14$\pm$0.25 &               - \\
         06\_49 &               - &  18.89$\pm$0.15 &   5.20$\pm$0.89 &     8.9$\pm$1.5 &   2.14$\pm$0.40 &  52.28$\pm$0.28 &  14.48$\pm$0.01 &               - \\
         06\_50 &               - &  32.13$\pm$0.01 &   2.32$\pm$0.96 &               - &   7.22$\pm$0.18 &    89.0$\pm$1.1 &  24.04$\pm$0.28 &               - \\
         06\_51 &               - &   3.66$\pm$0.72 &               - &               - &   1.32$\pm$0.79 &  18.33$\pm$0.04 &   4.68$\pm$0.56 &               - \\
         06\_53 &               - &               - &               - &               - &   1.02$\pm$0.11 &   3.13$\pm$0.10 &   0.79$\pm$0.28 &               - \\
  06\_54$^\dag$ &               - &     7.0$\pm$1.9 &               - &   2.62$\pm$0.63 &     2.8$\pm$1.8 &  11.69$\pm$0.03 &   3.32$\pm$0.13 &               - \\
         06\_55 &               - &    27.1$\pm$2.0 &     4.5$\pm$1.5 &   9.39$\pm$0.52 &   6.88$\pm$0.93 &  72.86$\pm$0.86 &  20.47$\pm$0.16 &   0.49$\pm$0.27 \\
  06\_56$^\dag$ &               - &     9.7$\pm$1.4 &     4.5$\pm$2.0 &   9.21$\pm$0.31 &   2.19$\pm$0.55 &  23.06$\pm$0.20 &   8.79$\pm$0.74 &               - \\
         06\_57 &               - &    20.7$\pm$3.7 &     1.9$\pm$1.3 &   6.12$\pm$0.31 &   4.18$\pm$0.22 &  72.02$\pm$0.46 &  14.54$\pm$0.06 &               - \\
         06\_59 &               - &  43.13$\pm$0.16 &     8.2$\pm$5.5 &    10.2$\pm$1.3 &   9.41$\pm$0.13 & 119.40$\pm$0.49 &  31.76$\pm$0.62 &     3.0$\pm$1.1 \\
         06\_61 &               - &    86.9$\pm$2.1 &     4.9$\pm$2.0 &    21.2$\pm$1.7 &  20.99$\pm$0.15 &   241.7$\pm$3.3 &  63.60$\pm$0.28 &   1.48$\pm$0.44 \\
         06\_62 &               - &  12.81$\pm$0.47 &   2.84$\pm$0.35 &   5.11$\pm$0.32 &   1.57$\pm$0.11 &  16.03$\pm$0.06 &   5.85$\pm$0.05 &               - \\
         06\_64 &               - &  31.74$\pm$0.40 &               - &               - &   6.02$\pm$0.37 &    88.0$\pm$1.6 &               - &   1.06$\pm$0.24 \\
         06\_65 &               - &  94.17$\pm$0.64 &     6.4$\pm$1.3 &  34.25$\pm$0.95 &  24.55$\pm$0.21 &   262.9$\pm$2.9 &  71.36$\pm$0.60 &   3.80$\pm$0.21 \\
         06\_68 &               - &    25.6$\pm$1.3 &               - &     5.8$\pm$2.3 &   4.36$\pm$0.39 &  71.20$\pm$0.62 &  16.08$\pm$0.04 &   0.55$\pm$0.11 \\
         06\_70 &               - &     5.9$\pm$3.0 &   3.52$\pm$0.95 &     6.5$\pm$1.2 &   1.16$\pm$0.49 &  13.27$\pm$0.18 &   3.63$\pm$0.04 &   0.24$\pm$0.24 \\
\hline \hline
\end{tabular}
\end{table*}

\begin{table*}
 \caption{Fluxes of the spectral lines observed in NGC6907 using the
  slit identified by the year of the observation and its ID. Slits flagged with a "$^\dag$" were removed
  from our analysis due the underlying Balmer absorption.}
 \label{tblflx-3}
  \centering
\begin{tabular}{crrrrrrrr}
\hline \hline
\multicolumn{ 9}{c}{{\bf NGC6907 - Fluxes [10$^{-17}$erg/cm$^{2}$/s/$\AA$]}} \\
\hline \hline
{\bf Slit} & \multicolumn{ 1}{c}{\bf [O~{\sc ii}]} & \multicolumn{ 1}{c}{\bf H$\beta$} & \multicolumn{ 1}{c}{\bf [O~{\sc iii}]} & \multicolumn{ 1}{c}{\bf [O~{\sc iii}]} & \multicolumn{ 1}{c}{\bf [N~{\sc ii}]} & \multicolumn{ 1}{c}{\bf H$\alpha$} & \multicolumn{ 1}{c}{\bf [N~{\sc ii}]} & \multicolumn{ 1}{c}{\bf [Ar~{\sc iii}]} \\
{\bf Year$\_$ID} & {\bf (3727$\AA$)} & {\bf (4861$\AA$)} & {\bf (4959$\AA$)} & {\bf (5007$\AA$)} & {\bf (6548$\AA$)} & {\bf (6563$\AA$)} & {\bf (6584$\AA$)} & {\bf (7136$\AA$)} \\
\hline
          05\_08 &      390$\pm$140 &    227.7$\pm$2.1 &     43.3$\pm$1.2 &  116.00$\pm$0.90 &                - &                - &                - &                - \\
   05\_10$^\dag$ &       123$\pm$52 &    110.7$\pm$2.1 &     46.0$\pm$3.1 &    107.4$\pm$2.0 &                - &                - &                - &                - \\
          05\_11 &       724$\pm$39 &    764.3$\pm$2.2 &    236.1$\pm$8.3 &    682.9$\pm$9.9 &                - &                - &                - &                - \\
          05\_12 &     3300$\pm$360 &   1963.0$\pm$3.5 &    360.6$\pm$5.3 &   1024.0$\pm$9.5 &                - &                - &                - &                - \\
          05\_14 &    306.2$\pm$6.6 &    541.4$\pm$2.7 &       105$\pm$38 &    301.6$\pm$3.3 &                - &                - &                - &                - \\
   05\_15$^\dag$ &                - &   1335.0$\pm$2.5 &    52.1$\pm$14.5 &    332.3$\pm$3.4 &                - &                - &                - &                - \\
          05\_16 &    9000$\pm$1300 &      8157$\pm$61 &       458$\pm$12 &   1351.0$\pm$7.5 &                - &                - &                - &                - \\
          05\_17 &      1476$\pm$97 &  610.10$\pm$0.45 &     14.8$\pm$4.9 &        82$\pm$11 &                - &                - &                - &                - \\
          05\_18 &                - &      2129$\pm$26 &    693.8$\pm$5.4 &      2031$\pm$19 &  548.00$\pm$0.75 &      4436$\pm$16 & 1620.00$\pm$0.50 &                - \\
          05\_19 &    3400$\pm$1100 &    13540$\pm$450 &    175.5$\pm$6.8 &     1130$\pm$220 &                - &                - &                - &                - \\
          05\_20 &       256$\pm$46 &    411.3$\pm$3.8 &        11$\pm$11 &     16.7$\pm$7.4 &                - &                - &                - &                - \\
          05\_21 &        89$\pm$39 &    347.7$\pm$2.8 &      7.7$\pm$3.8 &     23.2$\pm$2.0 &                - &                - &                - &                - \\
   05\_22$^\dag$ &       173$\pm$18 &    388.8$\pm$2.4 &     15.5$\pm$3.5 &     71.5$\pm$7.5 &                - &                - &                - &                - \\
          05\_23 &      380$\pm$120 &    804.3$\pm$2.7 &     33.7$\pm$9.2 &        64$\pm$17 &                - &                - &                - &                - \\
          05\_24 &        74$\pm$26 &     63.0$\pm$1.8 &     10.0$\pm$2.3 &     12.4$\pm$3.6 &                - &                - &                - &                - \\
          05\_25 &    2800$\pm$1200 &      1920$\pm$25 &  207.60$\pm$0.80 &    753.2$\pm$1.7 &                - &                - &                - &                - \\
          05\_26 &        99$\pm$28 &    266.8$\pm$1.0 &     30.6$\pm$2.5 &   91.60$\pm$0.46 &                - &                - &                - &                - \\
          05\_27 &     2860$\pm$660 &   1248.0$\pm$1.5 &    460.2$\pm$7.1 &   1309.0$\pm$9.5 &                - &                - &                - &                - \\
   05\_28$^\dag$ &                - &    4.97$\pm$0.47 &      3.0$\pm$1.5 &     15.5$\pm$3.7 &                - &                - &                - &                - \\
   05\_29$^\dag$ &        48$\pm$17 &     72.8$\pm$6.1 &     50.0$\pm$1.3 &  136.20$\pm$0.65 &                - &                - &                - &                - \\
   05\_31$^\dag$ &                - &   48.05$\pm$0.89 &     27.0$\pm$3.0 &   73.01$\pm$0.52 &                - &                - &                - &                - \\
   05\_33$^\dag$ &     13.4$\pm$5.0 &      4.0$\pm$2.0 &                - &      3.2$\pm$1.2 &                - &                - &                - &                - \\
   05\_34$^\dag$ &        50$\pm$36 &     11.9$\pm$1.8 &      3.4$\pm$1.6 &      4.9$\pm$2.7 &                - &                - &                - &                - \\
   05\_35$^\dag$ &        72$\pm$18 &     37.0$\pm$4.3 &     26.9$\pm$5.8 &     70.5$\pm$1.5 &                - &                - &                - &                - \\
   06\_01$^\dag$ &                - &     18.3$\pm$1.3 &                - &                - &      2.5$\pm$1.3 &   44.72$\pm$0.28 &   19.03$\pm$0.54 &                - \\
          06\_02 &                - &   12.25$\pm$0.69 &      9.8$\pm$1.1 &      8.6$\pm$1.0 &    2.76$\pm$0.18 &   30.31$\pm$0.16 &    9.54$\pm$0.02 &    0.58$\pm$0.15 \\
          06\_03 &                - &    194.2$\pm$7.2 &    260.3$\pm$4.0 &       749$\pm$16 &   14.88$\pm$0.11 &    439.2$\pm$1.7 &     54.1$\pm$2.1 &                - \\
          06\_05 &                - &   41.73$\pm$0.34 &   21.89$\pm$0.52 &   78.62$\pm$0.13 &   12.04$\pm$0.77 &  113.80$\pm$0.95 &   35.66$\pm$0.30 &    1.49$\pm$0.44 \\
          06\_07 &                - &  162.50$\pm$0.75 &        36$\pm$19 &   54.62$\pm$0.66 &   54.14$\pm$0.39 &  457.90$\pm$0.05 &  160.80$\pm$0.90 &      3.0$\pm$1.6 \\
          06\_08 &                - &      2017$\pm$45 &  253.30$\pm$0.30 &       785$\pm$19 &    426.7$\pm$9.3 &     4000$\pm$120 &      1325$\pm$40 &     34.3$\pm$5.0 \\
          06\_09 &                - &     53.9$\pm$4.1 &     12.6$\pm$4.2 &     39.5$\pm$4.9 &     35.3$\pm$3.6 &    285.5$\pm$2.4 &     97.8$\pm$2.2 &                - \\
   06\_10$^\dag$ &                - &   47.41$\pm$0.19 &    1.49$\pm$0.98 &      5.7$\pm$1.8 &   17.39$\pm$0.41 &  131.90$\pm$0.25 &   50.63$\pm$0.40 &    0.73$\pm$0.28 \\
          06\_12 &                - &   24.82$\pm$0.03 &     19.7$\pm$3.2 &   47.50$\pm$0.20 &    6.36$\pm$0.42 &   69.71$\pm$0.67 &   15.34$\pm$0.03 &                - \\
          06\_14 &                - &     31.3$\pm$4.9 &     24.0$\pm$2.4 &   16.90$\pm$0.25 &    1.68$\pm$0.18 &   39.14$\pm$0.18 &    8.90$\pm$0.72 &                - \\
          06\_15 &                - &    365.7$\pm$1.8 &  124.40$\pm$0.45 &    409.4$\pm$2.7 &   96.56$\pm$0.21 &   1021.0$\pm$4.5 &    300.4$\pm$1.1 &   18.00$\pm$0.05 \\
          06\_16 &                - &      3073$\pm$20 &    226.9$\pm$1.8 &       713$\pm$22 &   1049.0$\pm$1.5 &     8700$\pm$110 &      3307$\pm$38 &   69.50$\pm$0.44 \\
          06\_17 &                - &    172.1$\pm$3.3 &      9.9$\pm$2.7 &     18.6$\pm$3.6 &   46.96$\pm$0.33 &  488.10$\pm$0.35 &  164.20$\pm$0.15 &      2.4$\pm$1.0 \\
          06\_20 &                - &       409$\pm$12 &     77.6$\pm$5.0 &    210.9$\pm$3.2 &    123.3$\pm$1.1 &   1161.0$\pm$5.0 &    378.7$\pm$2.3 &                - \\
          06\_21 &                - &   2515.0$\pm$5.0 &    888.8$\pm$3.1 &   2664.0$\pm$5.0 &    821.7$\pm$7.4 &      7166$\pm$59 &   2320.0$\pm$2.0 &  131.70$\pm$0.20 \\
          06\_22 &                - &     46.3$\pm$1.1 &                - &     10.8$\pm$1.5 &     20.2$\pm$2.3 &  217.70$\pm$0.30 &   64.39$\pm$0.63 &                - \\
          06\_23 &                - &       2663$\pm$2 &    146.5$\pm$9.7 &  551.5$\pm$37.15 &      1014$\pm$64 &      7670$\pm$53 &      3107$\pm$17 &   49.78$\pm$0.74 \\
          06\_24 &                - &     3220$\pm$120 &    100.3$\pm$7.7 &       371$\pm$67 &      1428$\pm$30 &    10280$\pm$400 &     4620$\pm$250 &   17.31$\pm$0.58 \\
          06\_25 &                - &    338.0$\pm$3.7 &    117.2$\pm$0.8 &    345.5$\pm$3.5 &  119.70$\pm$0.20 &    935.9$\pm$4.3 &    365.9$\pm$3.2 &                - \\
          06\_26 &                - &    211.1$\pm$6.5 &     56.5$\pm$3.3 &       140$\pm$14 &   69.68$\pm$0.27 &    565.5$\pm$3.9 &    207.3$\pm$1.4 &      3.6$\pm$1.3 \\
          06\_27 &                - &     54.0$\pm$1.2 &     12.8$\pm$4.9 &     69.9$\pm$2.7 &    7.40$\pm$0.39 &  153.40$\pm$0.45 &   32.67$\pm$0.06 &    3.16$\pm$0.80 \\
          06\_28 &                - &   1006.0$\pm$0.5 &  271.60$\pm$0.85 &    852.8$\pm$2.0 &    271.2$\pm$3.1 &      2847$\pm$16 &    809.6$\pm$7.4 &   36.72$\pm$0.07 \\
          06\_29 &                - &    630.4$\pm$3.7 &    123.1$\pm$3.3 &    379.8$\pm$3.6 &    193.8$\pm$1.8 &   1787.0$\pm$6.5 &  596.00$\pm$0.45 &     19.4$\pm$3.2 \\
          06\_37 &                - &      9.1$\pm$2.8 &      4.7$\pm$1.3 &   20.12$\pm$0.37 &    0.54$\pm$0.05 &   16.64$\pm$0.08 &    3.66$\pm$0.17 &    0.18$\pm$0.11 \\
          06\_40 &                - &      5.2$\pm$1.7 &    4.94$\pm$0.45 &   20.59$\pm$0.84 &    1.30$\pm$0.17 &   13.57$\pm$0.21 &    3.87$\pm$0.44 &                - \\
          06\_45 &                - &     77.6$\pm$2.9 &     36.4$\pm$4.9 &   77.19$\pm$0.01 &   39.82$\pm$0.24 &    353.0$\pm$3.5 &  119.00$\pm$0.80 &    6.81$\pm$0.32 \\
          06\_46 &                - &     17.4$\pm$1.5 &                - &      9.2$\pm$3.0 &      3.9$\pm$1.4 &   85.11$\pm$0.93 &   28.40$\pm$0.77 &                - \\
          06\_47 &                - &   54.80$\pm$0.07 &   14.45$\pm$0.38 &   42.58$\pm$0.41 &   14.82$\pm$0.15 &    151.8$\pm$1.7 &   45.24$\pm$0.33 &      3.6$\pm$1.9 \\
          06\_48 &                - &    139.9$\pm$8.9 &      9.4$\pm$4.3 &                - &   34.16$\pm$0.87 &       309$\pm$20 &  131.30$\pm$0.01 &                - \\
   06\_49$^\dag$ &                - &    126.6$\pm$2.1 &                - &      7.2$\pm$2.6 &     59.3$\pm$2.2 &       501$\pm$24 &       192$\pm$18 &                - \\

\hline \hline
\end{tabular}
\end{table*}

\begin{table*}
 \contcaption{}
 \label{}
  \centering
\begin{tabular}{crrrrrrrr}
\hline \hline
\multicolumn{ 9}{c}{{\bf NGC6907 - Fluxes [10$^{-17}$erg/cm$^{2}$/s/$\AA$]}} \\
\hline \hline
{\bf Slit} & \multicolumn{ 1}{c}{\bf [O~{\sc ii}]} & \multicolumn{ 1}{c}{\bf H$\beta$} & \multicolumn{ 1}{c}{\bf [O~{\sc iii}]} & \multicolumn{ 1}{c}{\bf [O~{\sc iii}]} & \multicolumn{ 1}{c}{\bf [N~{\sc ii}]} & \multicolumn{ 1}{c}{\bf H$\alpha$} & \multicolumn{ 1}{c}{\bf [N~{\sc ii}]} & \multicolumn{ 1}{c}{\bf [Ar~{\sc iii}]} \\
{\bf Year$\_$ID} & {\bf (3727$\AA$)} & {\bf (4861$\AA$)} & {\bf (4959$\AA$)} & {\bf (5007$\AA$)} & {\bf (6548$\AA$)} & {\bf (6563$\AA$)} & {\bf (6584$\AA$)} & {\bf (7136$\AA$)} \\
\hline
          06\_50 &                - &     84.6$\pm$5.3 &      9.2$\pm$2.9 &      9.8$\pm$1.5 &     44.6$\pm$2.1 &       523$\pm$20 &       165$\pm$12 &                - \\
          06\_51 &                - &   80.41$\pm$0.76 &      9.7$\pm$3.4 &     44.8$\pm$3.5 &     39.0$\pm$1.8 &    306.3$\pm$9.5 &    124.1$\pm$1.6 &    1.35$\pm$0.35 \\
          06\_53 &                - &    194.3$\pm$2.5 &                - &                - &   26.82$\pm$0.65 &    239.3$\pm$9.4 &    101.4$\pm$3.7 &                - \\
   06\_54$^\dag$ &                - &    6.49$\pm$0.29 &                - &    3.09$\pm$0.40 &      2.3$\pm$1.5 &   12.23$\pm$0.83 &    5.79$\pm$0.71 &                - \\
          06\_55 &                - &        49$\pm$14 &     11.5$\pm$1.7 &     16.9$\pm$3.6 &   14.67$\pm$0.34 &     80.5$\pm$1.7 &   39.92$\pm$0.38 &                - \\
          06\_57 &                - &  244.80$\pm$0.40 &   42.48$\pm$0.65 &  145.40$\pm$0.60 &   74.15$\pm$0.33 &    685.0$\pm$6.0 &    229.2$\pm$2.4 &    9.18$\pm$0.31 \\
          06\_61 &                - &      7.3$\pm$2.1 &     23.0$\pm$3.4 &   69.91$\pm$0.13 &                - &    0.95$\pm$0.41 &    2.64$\pm$0.99 &                - \\
\hline \hline
\end{tabular}
\end{table*}

\end{document}